\documentclass[twocolumn]{aastex61}

\usepackage{savesym}
\savesymbol{tablenum}
\usepackage[separate-uncertainty=true]{siunitx}
\restoresymbol{SIX}{tablenum}
\usepackage{amsmath}
\usepackage{bm}
\usepackage{tikz}
\usetikzlibrary{calc}
\usepackage{todonotes}
\usepackage{color}

\renewcommand{\d}[2]{\frac{d#1}{d#2}}

\shorttitle{Formation of the Orion Fingers}

\begin{document}

\title{Formation of Orion Fingers}

\author{Ross Dempsey}
\affiliation{Department of Physics and Astronomy, Johns Hopkins University, 3400 N. Charles St., Baltimore, MD 21218, USA}
\author{Nadia L. Zakamska}
\affiliation{Department of Physics and Astronomy, Johns Hopkins University, 3400 N. Charles St., Baltimore, MD 21218, USA}
\author{James E. Owen}
\affiliation{Astrophysics Group, Imperial College London, Prince Consort Road, London SW7 2AZ, UK}

\begin{abstract}
``Orion fingers" are a system of dozens of bowshocks, with the wings of shocks pointing to a common system of origin, which is centered on a dynamically disintegrating system of several massive stars. The shock heads propagate with velocities of up to 300-400 km/s, but the formation and physical properties of the ``bullets'' leading the shocks are not known. Here we summarize two possible scenarios for the formation of the ``bullets'' and the resulting bowshocks (``fingers''). In the first scenario, bullets are self-gravitating, Jupiter-mass objects which were formed rapidly and then ejected during the strong dynamical interactions of massive stars and their disks. This scenario naturally explains the similar timescales for the outflow of bullets and for the dynamical interaction of the massive stars, but has some difficulty explaining the observed high velocities of the bullets. In the second scenario, bullets are formed via hydrodynamic instabilities in a massive, infrared-driven wind, naturally explaining the high velocities and the morphology of outflow, but the bullets are not required to be self-gravitating. The processes that created the Orion fingers are likely not unique to this particular star-forming region and may result in free-floating, high-velocity, core-less planets.
\end{abstract}


\section{Introduction} \label{sec:intro}

``Orion fingers'' in the Orion star-forming region were discovered via narrow-band imaging of near-infrared shock indicators such as ro-vibrational H$_2$ \citep{tayl84} and [FeII]$\lambda$1.644\micron\ \citep{alle93} emission lines. The morphology of the fingers -- narrow parabolae and cones with vertices pointing away from a common origin and arms pointing back toward it -- is rather similar to that of Herbig-Haro objects, commonly found in star-forming regions and usually powered by collimated outflows (jets) from young stars. However, the sheer number of fingers -- dozens -- combined with their similar propagation timescales and a seemingly common point of origin suggests a different provenance. The bullets may have been formed and ejected as part of a strong dynamical interaction between massive stars \citep{alle93}, or may have formed \emph{in situ} due to hydrodynamical instabilities in a stellar wind \citep{ston95}.

As seen in the plane of the sky, the fingers are grouped into two wide-angle fans, one pointing in the north-west direction and a less extensive one pointing to the south-east, producing an overall hourglass shape \citep{ball15}. Recent ALMA observations suggest that the outflow is intrinsically spherically symmetric \citep{ball17}, so the hourglass shape in H$_2$ and [Fe II] is likely a result of obscuration. Accurate proper motions \citep{ball11,ball15} and radial velocities \citep{niss12, youn16} have been measured for dozens of fingers. Their kinematic ages, obtained by dividing the projected distance from the presumed origin by the proper motion, range between 500 and 2000 years \citep{ball11, ball15}, and the maximum observed velocity in the plane of the sky reaches over \SI{300}{km/s}.

The fingers are observable due to bow shocks propagating through the interstellar medium (ISM). The leading tips of these bow shocks are seen most prominently in [FeII]$\lambda$1.644\micron, whereas the trailing shocks are seen better in H$_2$. This is consistent with a high-temperature shock head liberating FeII ions from dust and dissociating H$_2$, while the lower temperature wakes propagate into a dusty and molecular gas-rich medium. Some of the fingers also appear to be splitting apart \citep{ball15}, as if a bullet is fragmenting.

The nature of the bullets -- how they are confined, how massive they are and how they were ejected -- remains unknown. A key piece of observational evidence is that the fingers trace back to a small ($<1000$ AU) region \citep{ball11} where several massive run-away stars or tight binaries also originated 500 years ago \citep{rodr05, gome05, gome08, luhm17}, with few, if any, bullets having kinematic ages smaller than this value. This is consistent with bullets being ejected during a strong interaction between a handful of massive stars, which also led to the ejection of some or all of these stars \citep{ball05, chat12}. It is also consistent with bullets forming due to a hydrodynamical instability in wind from these massive stars \citep{ston95}, if launched at the same time as the stars were ejected.


In Section \ref{sec:bullets} we model a presumed dynamical interaction leading to the ejection of bullets, and we derive theoretical and observational constraints on the masses and sizes of these bullets. In Section \ref{sec:winds} we model an alternative scenario in which hydrodynamical instabilities in massive stellar winds led to the formation of the fingers. In Section \ref{sec:disc} we discuss the overall process from bullet formation to acceleration and its relevance to a variety of observations. We conclude in Section \ref{sec:conc}.

\section{Acceleration of bullets in stellar interaction}\label{sec:bullets}

The first scenario we consider is that of bullets formed and ejected during the stellar interaction which produced the runaway BN object, Source I, and source x. We envision a system of several massive stars with massive gas disks which start off as marginally gravitationally stable. As these stars begin their gravitational interaction, protoplanetary disks around them become gravitationally unstable and collapse into the compact objects which form the bullets. The gravitational instability proceeds on timescales similar to or shorter than the typical timescale of the stellar interaction, and during the last few close passages of the host stars, the formed bullets are ejected from the system to form the presently observed outflow. The hypothesis of the Orion fingers as by-products of ejected proto-planets was independently suggested by N.Z. Scoville (Bally, priv. comm.).

We assume throughout that these bullets are spherical with mass $M_b$ and radius $R_b$. In Section \ref{sec:simulations}, we present the results of $N$-body simulations of the ejection of the bullets. In Section \ref{sec:mass_size}, we derive several observational and dynamical constraints on $M_b$ and $R_b$ and show the narrow parameter space which the bullets must occupy in this scenario, and in Section \ref{sec:bullet_formation} we give a theoretical consideration of their confinement. Finally, in Section \ref{sec:bullet_feasibility}, we discuss the feasibility of this scenario in light of all these constraints.

\subsection{Velocities of bullets}
\label{sec:simulations}

If the Orion bullets were formed and ejected during a stellar interaction, then we expect their kinematics to be explained by the dynamics of this interaction. There are four objects in the BN/KL region which most likely participated in the interaction: the BN object, the likely binary Source I, and source x \citep{fari18}. It was originally suggested that the BN object was part of a triple system with $\theta^1$ Ori C, a binary system in the Trapezium cluster \citep{chat12}. However, this scenario requires a fine-tuned interaction between the BN star and the group of stars that produced I and x runaways, and  recent proper motion measurements have ruled out this history \citep{godd11}. \cite{ball05} proposed an alternative scenario in which the dynamical interaction of the BN object, Source I, and a third member -- originally thought to be source n, but now believed to be source x \citep{luhm17} -- led to the formation of a tight binary or even a stellar merger in the Source I, with the extracted binding energy fueling the runaway of BN, I and x. 

Stellar dynamical interactions in the BN/KL region were extensively studied with simulations by \citet{fari18}. These authors use \num{e7} $N$-body simulations of binary-binary interactions to calculate the probability of producing the observed outcome -- Source I as a tight binary or a merger with apparent velocity $\sim 15$ km/s, the BN object with velocity $\sim 30$ km/s and source x with velocity $\sim 55$ km/s. They find that the probability of the observed outcome is quite low,  and other tensions -- such as the size and the mass of Source I \citep{plam16} -- continue to linger in the dynamical decay model for the runaway stars. 

Our goal in this section is to determine whether the observed kinematics of bullets can be produced in a self-consistent way in the same interaction that produced the observed runaway stars. Following \citet{fari18}, we adopt $M_{\rm I1}=M_{\rm I2}=7M_{\odot}$ (roughly consistent with recent observations of a \SI{15}{M_\odot} rotation curve \citealt{gins18}), $M_{\rm BN}=10M_{\odot}$ and $M_{\rm x}=2.5M_{\odot}$ and conduct $\sim 10^3$ simulations using REBOUND \citep{rein12, rein15b}. We use the \texttt{whfast} integrator for speed in exploring the parameter space, although we have conducted some simulations with the more accurate \texttt{IAS15} integrator and found the results to be qualitatively the same. Again following \citet{fari18} we only consider binary-binary interactions (which these authors suggest are the most likely ones to have the high probability of close interaction in a 4-body system), and we consider all possible initial binary couplings with semi-major axes 10 AU and 30 AU and random distributions of orientations and orbital phases. Binaries approach each other at 5 or 3 km/s and interact, and we record their dynamical state $\sim 10^4$ years later. About 10 per cent of the interactions successfully result in a I=I$_1+$I$_2$ binary and BN and x singles -- we call these encounters ``qualitatively successful". However, in this subset of the simulations we observe a strong correlation between the runaway velocities of I and BN, and only a half-dozen interactions result in outcomes with $v_{\rm I}<15$ km/s while having $v_{\rm BN}$ and $v_{\rm x}>15$ km/s. We thus clarify why \citet{fari18}, who required a strong (within 2$\sigma$) agreement between the simulated and the observed velocities, find such a low probability of a successful simulation outcome: the outcomes with highly discrepant velocities of I and BN (as is observed) are quite unlikely. In addition, the actual velocity of source I in the OMC frame may be closer to \SI{10}{km/s} \citep{dzib17}, exacerbating the discrepancy.

\begin{figure}
    \centering
    \includegraphics[trim=10cm 9.5cm 1.7cm 1.7cm, clip=true, width=\linewidth]{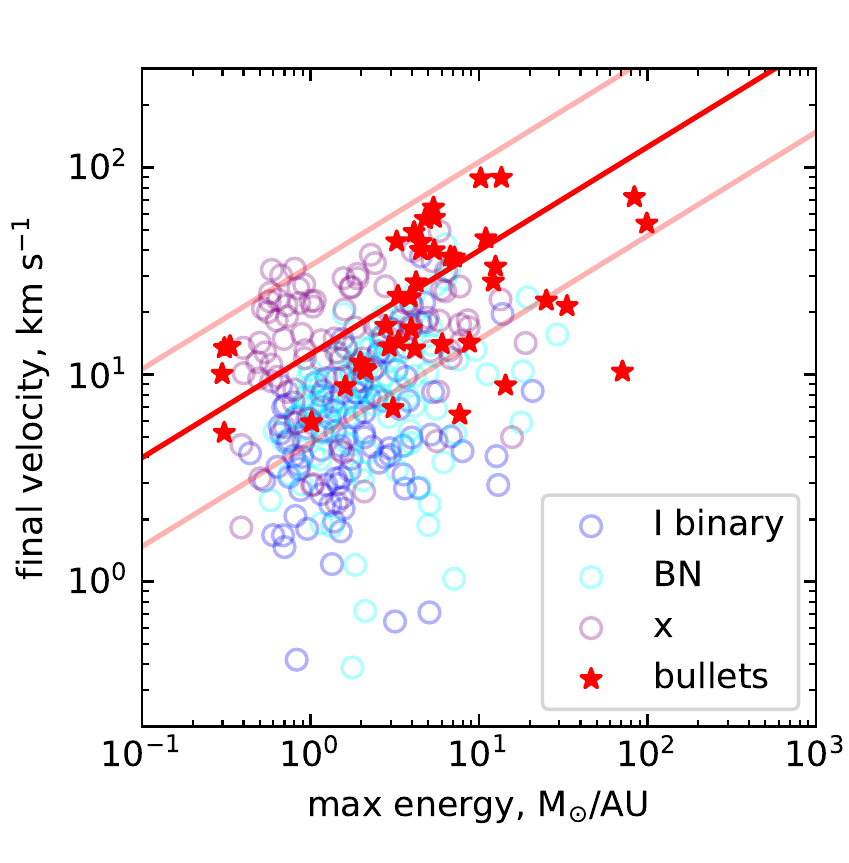}
    \caption{As a function of the maximal potential energy encountered during the interaction, in open circles we show terminal velocities of ejected stars in the ensemble of simulations that produce a qualitatively correct outcome (I$_1$+I$_2$ binary and BN and x singles). For one of these simulations, we also show (in filled stars) the terminal velocities of bullets as a function of their maximal encountered potential energy. The values are correlated within each subpopulation ($P_{\rm rank}=10^{-5}-10^{-2}$), and the $v_{\rm esc}\propto E_{\rm max}^{0.5}$ best fit for the bullets (and one population standard deviation around the fit) is shown with the solid line.} 
    \label{fig:esc_vel}
\end{figure}

Furthermore, for each of the escaping stars in the ensemble of the qualitatively successful simulations we find an anti-correlation between the final velocity and the distance of closest approach to one of the other stars during the simulations, or alternatively a positive correlation between the final velocity and the maximal binding energy reached during the interaction (the Spearman rank probability that these values are uncorrelated ranges between $10^{-2}$ and $10^{-5}$). These relationships are shown with circles for the qualitatively successful interactions in Figure \ref{fig:esc_vel}. For each escaping particle, its terminal velocity is weakly dependent on the maximal potential energy encountered during the simulation, with a scaling which is roughly consistent with the expected one, $v_{\rm esc}\propto E_{\rm max}^{1/2}$, albeit with a large scatter of $\sim 0.4$ dex. 

These relationships make it clear why it is difficult to achieve bullet velocities of $\ge 300$ km/s in the same interactions: while it is possible to accelerate the runaway stars to a few tens of km/s by passages within $\sim 1$ AU of other stars, the extremely high velocities of the bullets require passages within $\lesssim$ 0.1 AU of one of the stars. This means that even during an interaction that successfully produces the runaway stars, the cross-section of the requisite interaction for the bullets is much smaller, so only a few bullets may be fortuitously enough aligned during the interaction for such close encounters. Indeed, when we simulate some of our interactions that successfully produce $v_{\rm I}<15$ km/s while having $v_{\rm BN}$ and $v_{\rm x}>15$ km/s with test masses distributed in disks around the components of incoming binaries, we find that the velocity distribution does not exceed 100 km/s. The velocity distribution of the bullets in one such simulation is shown with stars in Figure \ref{fig:esc_vel}. A surface density distribution proportional to $R^{-3/2}$ \citep{weid77} between 0.1 and 10 AU is assumed in these simulations for the test particles. 

In conclusion, the scenario of the bullets being accelerated by the same dynamical interaction that produced the runaway stars suffers from some tension. First of all, the dynamical interaction that produced the runaway stars themselves is not yet well understood \citep{plam16, fari18}. Second, the observed velocities of the ejected stars suggest that they were accelerated in passages within $\sim 1$ AU of each other, but the 300 km/s velocities of the bullets cannot be easily produced by such encounters and likely require passages within 0.1 AU of one of the interacting stars, which may be true, but only for a small subset of bullets. 

Finally, it is not clear whether the sizes of the interacting stars are compact enough to enable such encounters. On the one hand, the Kelvin-Helmholtz timescale for massive stars is short enough that they should have contracted to their main-sequence size, $\ll 0.1$ AU \citep{kuip13}. On the other hand, high accretion rates on young massive stars can result in bloated photospheres with sizes of up to 1 AU. \citet{hoso09} hypothesize that BN is one such object. In addition, source I has a luminosity of $\sim\SI{1e4}{L_\odot}$ \citep{ment95} and a spectrum in reflected light corresponding to a temperature of $\sim\SI{4000}{K}$ \citep{test10}, which suggests a radius of
\begin{equation}\label{eq:size_srci}
R_\odot \left(\frac{T_\odot}{\SI{4000}{K}}\right)^4\left(\frac{\SI{1e4}{L_\odot}}{L_\odot}\right)^{1/2} \simeq \SI{1}{AU}.
\end{equation}  
However, the current accretion states and sizes of the runaway stars could well be different as a result of the interaction. In particular, source I may be the result of a merger, in which case the radii of I$_1$ and I$_2$ during the interaction could be very different from the present size \citep{fari18}. This leaves it unclear whether acceleration of bullets with such close passages is feasible. 

One caveat is that our simulations do not include any gas-dynamical effects. In principle, these could funnel the gas in the interacting disks toward their host stars and therefore establish a higher velocity scale for the escaping bullets. 

\subsection{Constraints on mass and size of bullets}\label{sec:mass_size}

Various physical and observational considerations allow us to place constraints on the mass and the size of the propagating bullets which generate the bow shocks. We work in units of Jupiter mass M$_{\rm J}$ and Jupiter radius R$_{\rm J}$. A number of observational and physical assumptions are used throughout our analysis. We assume that the distance to OMC-1 is $D=\SI{408}{pc}$ \citep{getm19}. We take the ambient sound speed to be \SI{2}{km/s}, consistent with molecular gas at a few hundred Kelvin; this is supported by the excitation temperature of CO in the region \citep{mcca97}. When computing dynamical constraints, we assume the fastest bullets are moving at \SI{300}{km/s} and that they formed \SI{500}{yr} ago \citep{ball15}.

Additionally, we take the ambient density to be \SI{e4}{cm^{-3}}, consistent with observations of larger structures in Orion \citep{ball87,ball15} as well as extinction measurements \citep{youn16} in the BN/KL \citep{beck67, klei67} region. While the actual density profile in the region is almost certainly not uniform, our calculations in this section concern the density encountered by a bullet over its \SI{\sim e5}{AU} trajectory. The kinematics will not be greatly influenced by a steep decline in density near the OMC-1 core, nor by small pockets of lower density throughout the region. Later on we will consider the density profile near the core in more detail.

Our first constraints are an upper bound on mass, due to the amount of material we expect to be available, and an upper bound on the size, based on the fact that Gemini adaptive optics cannot resolve the bullets themselves. There are several hundred shocks observed \citep{ball15}, and we expect each of these indicates a separate bullet which formed from a disk around one of the $\sim\SI{10}{M_\odot}$ stars. The disk should contain no more than about 10\% of the stellar mass to be gravitationally stable prior to the interaction, so we estimate a $\SI{\sim 1}{M_\odot}$ mass budget for the formation of bullets. This gives an upper bound on the individual bullet masses of
\begin{equation}\label{eq:mass_upper}
    M_b\lesssim \SI{3}{M_J}.
\end{equation} 
The observations of the outflow reached a resolution of \SI{.1}{"}, close to the \SI{0.07}{"} diffraction limit of the Gemini South Adaptive Optics Imager \citep{ball15}. Any object with $2R_b/D > \SI{.1}{"}$ should have been spatially resolved, which implies that 
\begin{equation}\label{eq:r_upper}
    R_b \lesssim \SI{8.4e4}{R_J}.
\end{equation}

The minimum mass of the bullets can be inferred from the line intensity of [Fe II] at \SI{1.644}{\micro\meter}, which is optically thin \citep{alle93}. Line intensities of $\SI{e-13}{erg/s/cm^2}$ suggest that $\num{\sim e46}$ Fe$^+$ ions are present in each of the bullets \citep{alle93}. Using the solar abundance of iron, we find a total minimal mass of $\SI{\sim e-3}{M_J}$ for the bullets. This estimate can be increased, since iron may be largely incorporated into dust grains not emitting [Fe II]. Fe II tends to be depleted by two orders of magnitude in the gas phase relative to the solar abundance \citep{harr84}; the shocks may liberate some additional iron, but we still expect at least an order of magnitude of depletion. Following this reasoning, \cite{alle93} suggest a mass bound \begin{equation}\label{eq:mass_lower}
    M_b\gtrsim \SI{e-2}{M_J}.
\end{equation}

We can learn more by considering the dynamics of the moving bullets. We denote the velocity of a bullet by $v_b$. The bullets are subject to ram pressure from the surrounding ISM, with density $\rho_\text{ISM}$. The force due to ram pressure is at least $\frac{1}{2}\rho_\text{ISM} v_b^2\times \pi R_b^2$, and can be greater if bow shocks alter the effective cross-sectional area \citep{shim85}. The bullets must reach their high velocities despite this deceleration. Using this minimum value for the force, the equation of motion implies
\begin{equation}
    v_b = \frac{v_0}{1+\frac{3}{8}\frac{\rho_\text{ISM}}{\rho_b}\frac{v_0\tau}{R_b}},
\end{equation}
where $\rho_b = M_b/(\frac{4}{3}\pi R_b^3)$ and $\tau$ is the time of flight. Solving for $v_0$,
\begin{equation}\label{eq:ram_pressure_v0}
    v_0 = \frac{v_b}{1-\frac{3}{8}\frac{\rho_\text{ISM}}{\rho_b}\frac{v_b\tau}{R_b}}.
\end{equation}
Thus, in order for the bullets to reach their present velocities despite ram pressure deceleration, $\rho_b R_b \sim M_b/R_b^2$ must exceed some critical value which is proportional to the velocity $v_b$. Substituting our assumed parameters for the highest velocity bullets ($\frac{\rho}{m_p} = \SI{e4}{cm^{-3}}$, $v_b = \SI{300}{km/s}$, and $\tau = \SI{500}{yr}$), this gives
\begin{equation}\label{eq:ram_deceleration}
    \frac{M_b}{R_b^2} \gtrsim \SI{3.2e-13}{M_J/R_J^2}.
\end{equation}


The velocity profile suggests that the bullets were ejected about 500 years ago, and they have traveled $\SI{\sim 0.15}{pc}$ in that time. Given this trajectory, momentum conservation requires that the bullets are $\sim 500$ times more dense than the surrounding ISM in order not to disintegrate \citep{ball15}. Assuming a background density of \SI{e4}{cm^{-3}}, this gives a minimum bullet density of
\begin{equation}\label{eq:density_bally}
    \frac{M_b}{(4\pi/3)R_b^3} \gtrsim \SI{1.6e-18}{M_J/R_J^3}.
\end{equation}

The survival of the bullets requires higher densities during the initial acceleration phase. In order to reach velocities of hundreds of kilometers per second, the bullets need to have close interactions with one or more of the stars, leading to large tidal forces on the bullets, which disrupt them unless they are of sufficient density. The tidal force experienced depends strongly on the minimal distance of stellar passage, which in turn can be inferred from the ejection velocity. The final velocities of the bullets should scale with the escape velocity $v\propto r^{-1/2}$ during the close interaction, while the tidal force scales as $F_t \propto r^{-3}$. It follows that the bullets must withstand a force $F_t \propto v^6$, so the large proper motions observed in the Orion outflow indicates that bullets would have had to withstand very large tidal forces. 

In order to fix the magnitude of the tidal forces in this relationship, we use results of the $N$-body simulations discussed in Section \ref{sec:simulations}. These simulations only produce bullets up to $\sim\SI{100}{km/s}$, which highlights a separate difficulty with the bullet scenario. We extrapolate the relationship between maximum tidal force and final velocity up to $\sim\SI{300}{km/s}$. Expressing the strength of the maximum tidal force as
\begin{equation}
    F_t = G\xi M_b R_b,
\end{equation}
we find
\begin{equation}\label{eq:tidal_fit}
    \log_{10}\left(\frac{v_b}{\SI{1}{km/s}}\right) = \left(2.54\pm 0.44\right) + \frac{1}{6}\log_{10}\left(\frac{\xi}{\SI{1}{M_J/R_J^3}}\right),
\end{equation}
where 0.44 dex represents one standard deviation in log-velocity. In order for self gravity to keep the bullets stable against the tidal force, we must have
\begin{equation}
    \frac{GM_b^2}{R_b^2} \gtrsim F_t \implies \frac{M_b}{R_b^3} \gtrsim \xi,
\end{equation}
so \eqref{eq:tidal_fit} gives a relationship between the bullet velocity distribution and the minimum density.

From the distribution of proper motions in \cite{ball15}, we find that the high-velocity bullets at $\sim \SI{300}{km/s}$ are about 1.7 standard deviations above the mean. This value leads to a minimum density of
\begin{equation}\label{eq:self_gravity}
    \frac{M_b}{(4\pi/3)R_b^3} \gtrsim \SI{2.1e-6}{M_J/R_J^3}.
\end{equation}


The bullets must condense out of material present in the surrounding circumstellar disk or molecular cloud. We consider the formation mechanism in more detail in Section \ref{sec:bullet_formation}, but the scenario is generally similar to the formation of a Jovian planet by gravitational instability \citep{boss97}. The bullets begin as low-density, low-temperature clumps of gas which slowly contract and heat. A key phase transition occurs when the core temperature reaches \SI{2500}{K} \citep{bode74}, at which point H$_2$ dissociates leading to rapid collapse. This transition is reached more quickly by more massive planets, but even for the largest masses allowed by the mass budget \eqref{eq:mass_upper} it would take on the order of \SI{e4}{yr} to reach this point. The dynamical timescale is only $\sim\SI{e3}{yr}$, and indeed only $\SI{e3}{yr}$ separate the onset of gravitational effects from the close interactions in the simulations discussed in Section \ref{sec:simulations}. Thus, the bullets have not yet reached core temperatures of \SI{2500}{K}, and are still contracting towards that point on Kelvin-Helmholtz timescales. Assuming they have virialized, we have
\begin{equation}
    M_b \approx \frac{2}{G}\left\langle u^2\right\rangle R_b = \frac{6kT}{G m_p} R_b,
\end{equation}
where $u$ is the thermal velocity of particles within the core. Since $T\le \SI{2500}{K}$, we have
\begin{equation}\label{eq:contraction}
    \frac{M_b}{R_b} \lesssim \SI{7e-2}{M_J/R_J}.
\end{equation}

The bullets are not detected as infrared point sources by Gemini \citep{ball15}, which places an upper bound on the surface temperature of the bullets. Using the detection limits for the 150 sec continuum exposures with the Gemini South Adaptive Optics Imager in the K$_s$ band, we find that objects in Orion with radii of $\SI{3}{R_J}$ would be undetected at surface temperatures $<1000$K and objects with radii of$\SI{10}{R_J}$ would be undetected at surface temperatures $<700$K, and at lower temperatures there are no meaningful constraints on size. In the scenario we are discussing here, the bullets are still collapsing and are in the beginning of the Kelvin-Helmholtz phase. Even though we cannot estimate their surface temperatures from first principles, we expect that their surfaces should be much cooler than those of young directly imaged exoplanets ($\la 1000$K at the peak surface temperature of their evolution; \citealt{maro08}), so this does not lead to a separate constraint.

The bullets must be large enough to power the wide bow shocks seen in Gemini images \citep{ball15}. In an oblique shock, the pressure discontinuity scales as \citep{whit74}
\begin{equation}
    \frac{\Delta P}{P} = M^{1/4} \left(\frac{r}{R_b}\right)^{-3/4},
\end{equation}
where $M$ is the Mach number and $r$ is the perpendicular distance to the bullet trajectory. We argue in Section \ref{sec:angles} that the pressure must be dominated by magnetic fields, with $P_\text{magnetic} \sim 200P_\text{gas}$. The magnetic fields do not dissipate energy, so $\Delta P = \Delta P_\text{gas}$. Using $P\approx P_\text{magnetic}$, we find
\begin{equation}
    \frac{\Delta P_\text{gas}}{P_\text{gas}} = \frac{P_\text{magnetic}}{P_\text{gas}} M^{1/4}\left(\frac{r}{R_b}\right)^{-3/4}.
\end{equation}
For a strong shock, we should have $\frac{\Delta P_\text{gas}}{P_\text{gas}} \ga 1$. Since the visible bow shocks stretch over \SI{1000}{AU} from the shock heads, this gives
\begin{equation}
    R_b \ga \SI{100}{R_J}.
\end{equation}



These constraints are all shown in the left panel of Figure \ref{fig:mr_diagram}. Together they eliminate all feasible phase space for bullets which could be the progenitors of the Orion fingers, provided we assume the bullets are confined by self-gravity during their acceleration. In the following section, we consider possible formation mechanisms for the bullets and discuss the possibility that they are confined by ram pressure instead.

Another potential constraint on the size of the shock head comes from the emission of the hot post-shock gas. By overlaying the deep X-ray image by \citet{gros05} on the Gemini image \citep{ball15}, we find no obvious point X-ray sources associated with the fingertips or diffuse X-ray emission in the wakes. In principle, with the post-shock temperature expected to be in the $10^5-10^6$K range, this non-detection sets an upper limit on the size of the propagating bullet. Indeed, X-ray emission with characteristic temperatures in this range is detected from the shock heads of a nearby Herbig-Haro object HH210 \citep{gros06}. Unfortunately, fingertips are embedded in a region with a much higher intervening column density ($N_{\rm H}\simeq 10^{22}-10^{24}$ cm$^{-2}$; \citealt{gros05}) than that seen in HH210, and that strongly hinders the detectability of soft X-rays. Given the uncertainties in the post-shock temperatures and the intervening column densities, we cannot place a strong constraint on the size of the bullet from the current non-detection of X-rays for any of the scenarios we discuss.

\subsection{Formation of confined bullets}\label{sec:bullet_formation}

\begin{figure*}
    \centering
    \includegraphics[width=\linewidth]{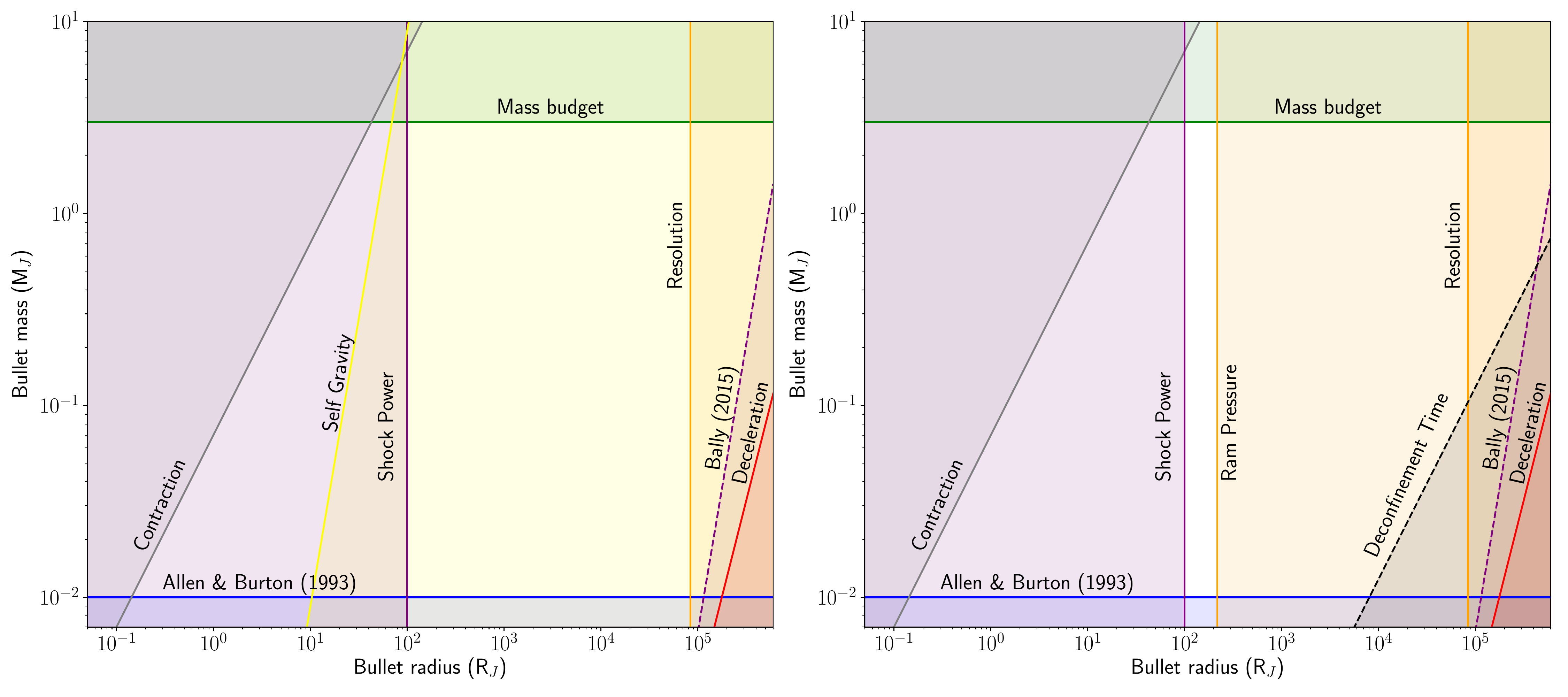}
    \caption{The various constraints on the mass and size of propagating bullets which may be the responsible for the formation of the Orion fingers. On the left we show the constraints assuming the bullets are self-gravitating during their acceleration, and on the right we show the constraints assuming the confinement is due to ram pressure. These constraints are described explicitly in Section \ref{sec:mass_size}, except for the ram pressure stability and confinement criteria, which are derived in Section \ref{sec:bullet_formation}. The density contrast constraint \eqref{eq:density_bally} from \cite{ball15} is shown as a dashed line, since the morphology suggests that some bullets might in fact be disintegrating, although we present an alternative explanation for these shapes in Section \ref{sec:disintegration}.}
    \label{fig:mr_diagram}
\end{figure*}

If the hundreds of shocks \citep{ball15} observed in Orion come from bullets, then either these bullets were pre-existing in disks around the stars before their dynamical interaction or they formed as a result of the dynamical interaction. In the former case, the bullets orbiting as planetesimals would have to be separated by several Hill radii to be dynamically stable \citep{fang13}. Given the bullet mass suggested by the constraints in Figure \ref{fig:mr_diagram}, dynamical stability implies that no more than a few tens of bullets could have pre-existed as planetesimals, so most of the bullets formed during the dynamical interaction. In this section we review possibilities for how the bullets could have formed.

If the bullets are gravitationally bound, then they most likely formed from material which was originally bound to the stars in disks. The disks would have suffered tidal disruptions during the dynamical interaction of the stars. In galaxy mergers, such disruptions lead to gaseous inflows which trigger nuclear starburst \citep{miho96}. In Orion, we have essentially the same process on a smaller scale: the dynamical interaction of protoplanetary disks leading to inflow which triggers a collapse into a population of self-gravitating bullets.

Starburst events in galaxy mergers are understood to be a consequence of angular momentum transport caused by a disruption of the axisymmetric galactic potential by its merger partner. This dynamical process is initially insensitive to the nature of the objects formed via the resulting gravitational instabilities, so we can use the scaling relations for a self-gravitating system to relate collisions of planetary disks to galaxy mergers. \cite{miho96} use galaxies of mass \SI{5.6e10}{M_\odot} and radius \SI{3.5}{kpc}, and find a burst in star formation rate about \SI{750}{Myr} after their collision. Thus, according to
\begin{equation}
    m\mapsto \alpha m,\qquad r\mapsto \beta r,\qquad t\mapsto \alpha^{-1/2}\beta^{3/2}t,
\end{equation}
we expect two protoplanetary disks of radius $\sim\!\SI{10}{AU}$ around stars of mass $\sim\!\SI{10}{M_\odot}$ to form dense clumps of gas after $\sim\!\SI{100}{yr}$.

After this time, the formation of the clumps is sensitive to the details of the feedback, which is different in protoplanetary and galactic disks. In galactic disks, stellar winds remove the gas suppressing star formation. In protoplanetary disks, the lack of such feedback should only enhance the efficiency of the process by allowing the unstable central region of the merger to continue to undergo gravitational collapse, producing more clumps \citep{armi99}. Additionally, the clumps can undergo Bondi-Hoyle accretion to grow to planet-scale masses.


A key constraint on this process is the stability of the clumps against the tidal forces of the orbiting protostars. If the clumps are to be accelerated to several hundred km/s and become the presently observed Orion bullets, then they must have a close interaction with the stars. If the bullets are not dense enough, they pass within the Roche limit and are tidally disrupted. 

This constraint is avoided if the bullets are not confined by self-gravity. Since the bullets are propagating through dense ISM, they could be confined by ram pressure instead. Ram pressure confinement has been studied in the case of high velocity jets from radio sources propagating through a relatively dense intergalactic medium \citep{bege89,loke92}. The accepted mechanism, which is also applicable in the regime we are concerned with, consists of three phases within a bow shock: the propagating material itself, a cocoon of shocked material around it, and then a shell of shocked ambient medium (IGM in the case of galactic jets, or ISM in our case). The cocoon serves as a reservoir of energy dissipated from the propagating material, and its thermal pressure becomes higher than the ambient medium \citep{bege89}. This allows for confinement of the propagating material, while also causing the cocoon to expand into the ambient medium. After sufficient time, this expansion leads to pressure balance, and confinement is lost \citep{loke92}.

If ram pressure is responsible for the confinement of the bullets, it must have kept them stable against the tidal forces they experienced during the stellar interaction. In addition, ram pressure could be the present confinement mechanism, as an alternative to the density contrast constraint \eqref{eq:density_bally}. We start by examining the latter possibility, and then apply our analysis to the tidal forces during the close-pass interaction. 

If the Orion bullets are to be presently confined by ram pressure, the time $\tau$ necessary for losing this pressure balance must not yet have elapsed. Following \cite{loke92}, we can compute this time. The bullet itself, with velocity $v_b$ and size $R_b$, will drive a bow shock as it moves through the ISM. Since the bullet is much denser than the medium, the shock head moves with a velocity approximately equal to $v_b$. There will also be a cocoon around the bullet trajectory, with radius $R_c$.

The volume of the cocoon is $\pi R_c^2 \int_0^{\tau} v_h\,dt$. If we assume the kinetic power $L_k = \pi R_b^2 \rho_b v_b^3$ is deposited with efficiency $\epsilon$ into the cocoon, then the pressure in the cocoon is given by
\begin{equation}\label{eq:cocoon_pressure}
    P_c = \frac{\epsilon \int_0^{\tau} L_k\,dt}{\pi R_c^2 \int_0^{\tau} v_h\,dt} \sim \frac{\epsilon L_k}{\pi R_c^2 v_h},
\end{equation}
where in the second expression we neglect the time dependence of $L_k$ and $v_h$. This pressure must be balanced with the ambient ram pressure, given by $\rho_\text{ISM}(R_c/\tau)^2$. Since we are concerned about deconfinement at late times, the relevant density ($\rho_{\rm ISM}/m_p=10^4$ cm$^{-3}$) is the density on large scales. This implies
\begin{equation}
    R_c = \epsilon^{1/4} \left(\frac{\rho_b}{\rho_\text{ISM}}\right)^{1/4}(v_b R_b)^{1/2} \tau^{1/2}.
\end{equation}
Since $R_c\propto \tau^{1/2}$, the cocoon pressure decreases as $P_c \propto \tau^{-1}$, and thus eventually ceases to be overpressured. This sets a time scale for deconfinement of the bullet. Given the sound speed $c_\text{ISM}$ in the surrounding medium and its adiabatic index $\gamma$, its pressure can be expressed as $P_\text{ISM} = \rho_\text{ISM} c_\text{ISM}^2/\gamma$. Setting this equal to $P_c$ gives a deconfinement time of
\begin{equation}\label{eq:deconfinement_time}
    \tau_\text{dc} \approx \sqrt{\epsilon}\gamma\sqrt{\frac{\rho_b}{\rho_\text{ISM}}}\frac{v_b R_b}{c^2_\text{ISM}}.
\end{equation}
We need $\tau_\text{dc} > \SI{500}{yr}$. Letting $\epsilon = 0.1$, $\gamma = \frac{7}{5}$, and $c_\text{ISM} = \SI{2}{km/s}$, this constraint amounts to
\begin{equation}
    \frac{M_b v_b^2}{R_b} \ge \SI{3.1e-3}{M_J (km/s)^2/R_J}
\end{equation}

The lowest velocity objects in the OMC-1 outflow have $v_b\sim\SI{50}{km/s}$ \citep{ball15}. Using this value, we find
\begin{equation}
    \frac{M_b}{R_b} \gtrsim \SI{1.2e-2}{M_J/R_J},
\end{equation}
which is shown as a dashed line in Figure \ref{fig:mr_diagram}, indicating that it need not apply if the bullets are self-gravitating.

In addition to this criterion for the present ram pressure confinement of the bullets, the ram pressure would have to be sufficient to stabilize the bullets against the tidal forces they experience during the initial acceleration stage. In Section \ref{sec:mass_size}, we showed that the bullets experience a maximum tidal force $F_t = G\xi M_bR_b$ where $\xi=\SI{8.9e-6}{M_J/R_J^3}$. From \eqref{eq:cocoon_pressure}, we expect that in the early stages of the bow shock, when $R_b\sim R_c$, the ram pressure leads to a force
\begin{equation}
    4\pi R_b^2 P_c \sim \frac{3\epsilon M_b v_b^2}{R_b}
\end{equation}
on the bullet. For this to exceed the tidal force, we must have
\begin{equation}
    R_b \lesssim \sqrt{\frac{3\epsilon}{G\xi}}v_b.
\end{equation}
Substituting $\epsilon = 0.1$, $v_b=\SI{50}{km/s}$, and $\xi=\SI{8.9e-6}{M_J/R_J^3}$, we find
\begin{equation}\label{eq:ram_upper}
    R_b \lesssim \SI{220}{R_J}.
\end{equation}
This gives a somewhat weaker constraint than if the bullets are confined by self-gravity against the same tidal forces. Nonetheless, it still leaves only a small range of feasible bullet radii, as shown in the right panel of Figure \ref{fig:mr_diagram}.

\subsection{Feasibility}\label{sec:bullet_feasibility}

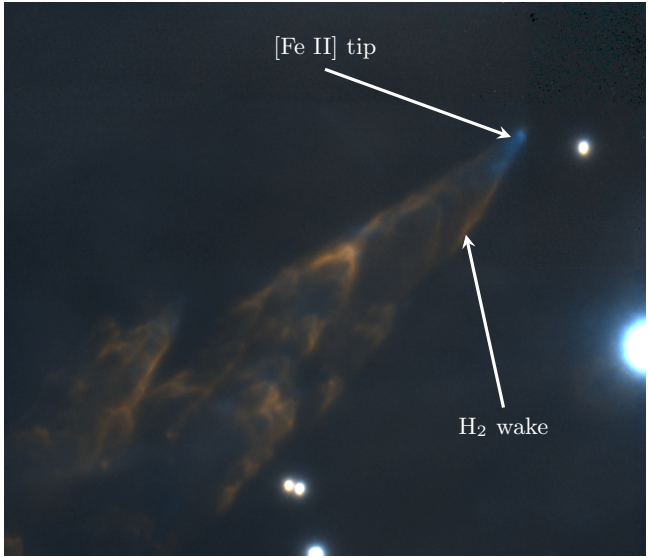
\begin{figure}
    \centering
    \begin{tikzpicture}
        \node at (0,0) (image) {\includegraphics[width=\linewidth]{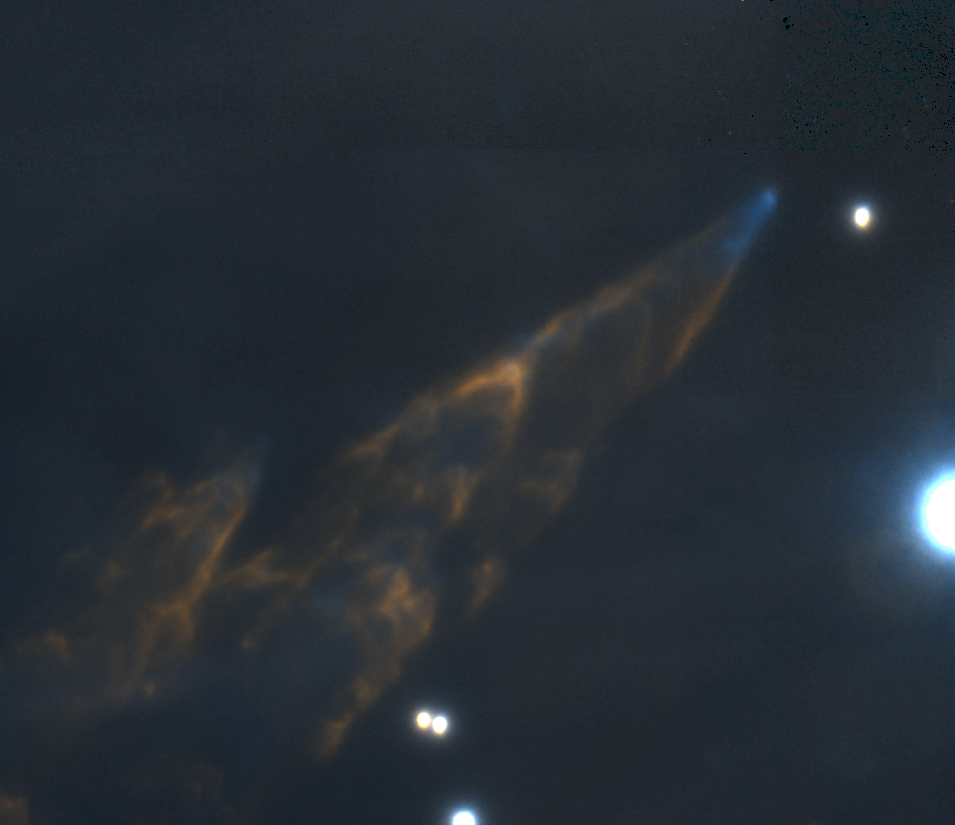}};
        \coordinate (tip) at ($(image.north east)+(-1.9,-1.9)$);
        \coordinate (wake) at ($(image.north east)+(-2.5,-3.2)$);
        \draw[-stealth,white,very thick] ($(image.north)+(0,-1)$) node[above] {[Fe II] tip} -- (tip);
        \draw[-stealth,white,very thick] ($(wake)+(0.5,-2.3)$) node[below] {H$_2$ wake} -- (wake);
    \end{tikzpicture}
    \caption{An example of the Orion bullets imaged in \citet{ball15}, showing [Fe II] in cyan and H$_2$ in orange. Reprinted with permission from Figure 1 of ``The Orion fingers: Near-IR adaptive optics imaging of an explosive protostellar outflow'' \citep{ball15}.}
    \label{fig:bullet_example}
\end{figure}

The formation of the Orion fingers by dynamically expelled bullets explains several observations. First among these is the morphology of the outflow. The conical structures pointing outwards from the center point can be readily understood as bow shocks formed by the bullets. This also explains the structure of their emission lines. [Fe II] is observed in the heads \citep{ball15}, indicating the presence of gaseous iron which has been liberated from dust grains. This is to be expected if dense bullets are plowing through ISM, causing dust destruction by either sputtering or heating. Furthermore, molecular hydrogen lines are observed in the wakes of the shock, and the timescale for reformation of molecular hydrogen after dissociation is long enough that the $\text{H}_2$ must have been present before the arrival of the bullet \citep{gold07}. The $\text{H}_2$ could survive in the wakes since they lie well outside the trajectory of the bullet, and thus were not dissociated by the head of the shock. Indeed, as we trace the bowshock from the head to the wings, it becomes more oblique, indicating a decrease in the effective shock velocity \citep{ball06}, and furthermore the effective Mach number of such shocks declines as $\propto d^{-3/4}$ away from the trajectory of the bullet \citep{whit74}. These milder postshock conditions do not lead to the dissociation of the molecules on the shock wings, so the $\text{H}_2$ is observed intact.

In addition to the [Fe II] and $\text{H}_2$ lines observed in \cite{ball15}, \cite{ball17} presented images of ``streamers'' in CO lines. The CO streamers point towards the $\text{H}_2$ fingers, but they have lower velocities and are not as far from the ejection source. These streamers can be explained by fast-moving bullets if we take projection accounts into consideration. We expect there to be some central column around the bullet which is shock-heated enough to dissociate CO, and the formation timescales for CO are too long \citep{glov10} to permit reformation within $\sim 500$ years. In some wider column, CO molecules can be excited by the propagation of the oblique shock away from the bullet. This will lead to a cylindrical shell of excited CO, which when projected onto the plane of the sky produces the observed streamers.

Additionally, in this scenario there is a clear reason why the minimal kinematic age of the fingers agrees closely with the time of the stellar dynamical interaction. The bullets are accelerated by multi-body interactions with the orbiting stars, so they are expelled from the center point during the same interaction as the one that leads to the runaway stars, and no later than the last closest approach that led to the disintegration of the stellar system. Indeed, while there is some spread of apparent kinematic ages between 500 and 1000 years \citep{ball11}, which may result from a combination of measurement uncertainties, deceleration and a true spread in ages, there are essentially no ``fingers" younger than 500 years. 

However, there are two serious obstructions to viewing the Orion fingers as bullets which were accelerated from a central point. The first of these is the difficulty in accelerating the bullets to the high observed velocities, \SIrange{300}{400}{km/s}. The second problem is the narrow parameter space allowed by the various physical constraints. While we presented a qualitative scenario for forming self-gravitating bullets in the relevant amount of time (the timescale of the dynamical interaction, $10^3$ years), the bullets would have to be too small to power shocks spanning $\sim\!\SI{1000}{AU}$. Alternatively, if the bullets are confined by ram pressure during the dynamical interaction, our analytic estimates suggest a very narrow range of feasible radii, $100 \la R_b/R_J \la 220$. The actual ability of bullets in this range of radii to survive by ram pressure confinement and power large shocks has not yet been verified by direct numerical simulations. 

While neither of these concerns is completely decisive, owing to the approximate nature of our calculations, they cast serious doubt on the dynamical ejection scenario, given that similar outflow morphologies have been observed in other star-forming regions \citep{saha08, smit09, liu13, zapa13, saha17}. This suggests that the formation mechanism should not require excessive fine-tuning.


One caveat of our analysis is that our simulations in Sec. \ref{sec:simulations} do not include any gas-dynamical effects. It follows that we are not sensitive to a potential mechanism for promoting the inward migration, and subsequent close-pass interactions, of the bullets. As the gas disks around the interacting stars collide, angular momentum transfer between the bullets and the gas in the presence of the non-axisymmetric potential can shift them towards radial orbits. If this occurred with self-gravitating bullets, then they could potentially be accelerated more efficiently than our simulations suggest while also remaining stable against the high tidal forces. Alternatively, the bullets could have formed from ram pressure confinement of gas which was accelerated during the interaction. There would then be no need for compact objects to survive large tidal forces. We leave the detailed analysis of these scenarios to future work; from our present analysis, we cannot fully rule out the formation of the Orion fingers by dynamical ejection of bullets.

\section{Formation of fingers in situ}\label{sec:winds}

The Orion bullets are likely related to the stellar interaction which occurred some 500 years ago. However, it is possible that the bullets did not form exactly at this time, but rather condensed out of a stellar wind which was launched during the interaction \citep{ston95}. We now explore this scenario, and show that it naturally resolves the primary concerns with the dynamical ejection of bullets during the interaction. 

In Section \ref{sec:hydro_formation}, we discuss the Rayleigh-Taylor instability in stellar winds. In Section \ref{sec:wind_dynamics}, we show that a massive stellar wind driven by radition pressure of infrared light from the stars can reach the velocities of several hundred km/s needed to explain the high-velocity Orion bullets. In Section \ref{sec:mass_size_wind} we discuss numerical simulations of this mechanism, and in Section \ref{sec:wind_feasibility} we discuss the feasibility of this scenario.

\subsection{Hydrodynamical formation of fingers}\label{sec:hydro_formation}

Massive stellar winds are naturally capable of generating velocities of several hundred \si{km/s} \citep{cast75}. Outflows have been identified around the BN source \citep{bunn95} and Source I \citep{hiro17}, so these high wind speeds are available in the vicinity of the Orion fingers. If a hydrodynamical instability in an outflow could trigger the formation of compact objects which create bow shocks as they continue to plow through the ISM, then we could naturally explain the kinematics and morphology of the Orion fingers \citep{ston95}.

In fact, we expect that rapid injection of energy into an inhomogeneous damping medium will generically produce the finger morphology seen in Orion. Any dense clumps which form in a wind or other outflow will experience less ram pressure deceleration, and will thus achieve higher speeds and propagate farther than the other material. These dense clumps and their wakes will then be visible as fingers stretching out from the center of the system. Such process has been directly observed by detonating an explosive in the center of a bed of dense particles \citep{fros12} or in a spherical shell of liquid \citep{miln17}. The explosion consists of particle jets which move more quickly than a trailing spherical shock wave \citep{fros12}.

Thus, if a gaseous outflow in the Orion system fragmented, the resulting dense clumps could be the progenitors of the present fingers. A key mechanism which can drive the fragmentation of an outflow is the Rayleigh-Taylor (R-T) instability. This instability is well-known in supernova remnants, where it can develop at the interface between the supernova blast wave and the ambient medium. In our scenario, we expect the instability is generated at the interface of a stellar wind emanating from the interacting young stars and the surrounding interstellar medium. The morphology of the Orion fingers, with the fingers propagating into the ambient medium ahead of the stellar wind, requires a somewhat different acceleration profile than what we see in supernova remnants. Following \cite{ston95}, we propose that the collision of a fast wind with a previously launched, slower wind could be responsible for the hydrodynamical instability which generated the Orion fingers.

\cite{ston95} used hydrodynamical simulations to confirm the morphological similarity between filaments due to an R-T instability in interacting stellar winds and the observed Orion fingers. The fast inner wind is decelerated by the outer shell, and fragments into small dense clumps which continue to propagate into the surrounding ISM. These clumps then form the bow shocks and Mach cones which we observe as the Orion fingers. We confirm this qualitative behavior in Section \ref{sec:mass_size_wind} with a modified version of the \cite{ston95} simulations.

The R-T instability can also be present in an astrophysical outflow as a result of radiation pressure which induces the requisite acceleration profile \citep{krum12}. We are not exploring this mechanism, instead focusing on the rapid deceleration due to a collision of winds which is more likely to produce phenomena on larger scales.

\subsection{Velocities of fingers}\label{sec:wind_dynamics}

There are two velocity scales present in the OMC-1 outflow. The median velocity is about $\SI{20}{km/s}$, and this velocity characterizes the $\SI{\sim 8}{M_\odot}$ of gas which makes up the bulk of the outflow. This component of the outflow is due to the gas that was gravitationally bound to the massive stars during the interaction. Since then, the massive stars have left the center of potential. If we take the gas to have an initial median distance of \SIrange{10}{20}{AU} from the BN source, which has a mass of $\SI{\sim 10}{M_\odot}$, then the resulting velocity now that this gas is freely expanding is $\SI{\sim 20}{km/s}$, as observed.

In contrast, the fingers have a range of significantly higher velocities. Gemini images of the Orion fingers with adaptive optics have enabled accurate measurement of proper motions of several fingertips \citep{ball15}. The velocities are \SIrange{200}{300}{km/s}, with one high velocity clump reaching $\SI{300(20)}{km/s}$. If the fingers formed via hydrodynamic instabilities, then their velocities must reflect the velocity of the wind from which they arose.

A high-velocity wind could potentially have formed due to the sudden release of gravitational binding energy during the interaction which led to the tight binary or stellar merger now observed as source I. This possibility has been previously considered for the Orion system \citep{ball11}, and more generally in the context of common envelope evolution \citep{ivan13}. Such a wind, upon colliding with the surrounding ISM, would be likely to trigger Rayleigh-Taylor instabilities similar to those seen in \citet{ston95}. This is an interesting possibility, but theoretically it suffers from a lack of understanding of the coupling mechanism between the gravitational binding energy and the resulting wind. In this scenario, a lot of material would need to originate close to the photospheres of the stars to get to the high observed velocities. 

In order to present a detailed analysis of the achievable wind speeds, we choose to instead focus on winds driven radiatively by the interacting stars. Stellar winds are well-studied and it is clear how radiative output couples with a wind to drive it to potentially very high velocities. Our choice to focus on this case does not preclude the possibility that the Orion fingers are driven by energy released as a result of the interaction through a less well-understood process.

The stars are very luminous but are enshrouded in dust, so we primarily consider IR radiation as the driving force for the wind. The dynamics of IR radiation-driven winds is still subject to many uncertainties \citep{krum12}. Here we follow \citet{salp74,thom15} for an approximate description of the dynamics of such a wind:
\begin{equation}\label{eq:wind_dynamics}
    v\d{v}{r} = -\frac{GM}{r^2} + \frac{\kappa_{\rm IR} L}{4\pi c r^2} - v^2 \frac{1}{M_\text{sh}}\d{M_\text{sh}}{r},
\end{equation}
where $M$ is the mass of the central driving source, $L$ is its luminosity, $\kappa_\text{IR}$ is the infrared opacity of the driven wind, and $M_\text{sh}(r)$ is the mass of the shell when it reaches a distance $r$. If we make the simplifying assumption that the mass is all swept up at once, so that $M_\text{sh}(r)$ is a step function, then the terminal velocity is \citep{salp74}
\begin{equation}\label{eq:v_terminal}
    v_\infty^2 = v_\text{esc}^2(r_0)(\Gamma-1),
\end{equation}
where $\Gamma = L/L_\text{Edd} = \kappa_{\rm IR} L/(4\pi GMc)$ is the Eddington ratio with respect to infrared opacity. The relationship between escape velocity and terminal velocity has been studied observationally in hot sources including O and B stars, and a strong linear relationship of $v_\infty \approx 3v_\text{esc}$ has been found \citep{abbo78}.

The BN object has a mass $\SI{\sim 10}{M_\odot}$ and a radius \SIrange{3.4}{4.3}{R_\odot} \citep{tan04}, so the escape velocity from its surface is approximately $\SI{600}{km/s}$. Its luminosity is a few times \SI{e4}{L_\odot} \citep{scov83}, which gives it an Eddington ratio up to $\Gamma\sim 1.5$. Equation (\ref{eq:v_terminal}) then gives a terminal velocity $v_\infty = \text{\SIrange{300}{400}{km/s}}$. In the simulations discussed in the following section, we find that the velocity of the fingers is inherited from the velocity of the fast wind, so this terminal velocity can explain the present motion of the fingers at hundreds of kilometers per second. Thus, in contrast to the explosive scenario, stellar winds are naturally capable of developing the several hundred km/s velocities necessary to explain the proper motions of the Orion fingers.

One potential concern with this mechanism is that it naturally generates a population of objects moving with the same velocity, whereas the observed outflow has a range of velocities with a Hubble-type relation between distance and velocity \citep{ball11}. Furthermore, the CO streamers associated with the outflow show a close correlation between projected distance and radial velocity \citep{ball17}. This can be explained by corrections to the thin spherical shell model of \cite{thom15}. If the optical depth of the wind varies across the shell prior to the wind launch, then different parts of the wind would be driven to different velocities, naturally producing a Hubble-type flow.



Typically, this model is made more complex by the presence of dust. Due to spherical expansion, the column density of dust in the wind will decrease with time, and the dust will become optically thin to IR photons before it becomes optically thin to UV photons. Between these two timescales, there is a phase in which incoming UV photons exert outward radiation pressure while the re-radiated IR photons have no opposite effect, which in principle could allow the wind to be driven to higher velocities \citep{thom15}. However, the sources present in the BN/KL outflow are still enshrouded by dust, so this UV-driving mechanism is likely unimportant.


The key question regarding the winds is whether IR radiation pressure can produce the velocities of hundreds of kilometers per second required to explain the observed proper motion of the Orion fingers. In \eqref{eq:wind_dynamics}, we see that mass-loading impedes the radiation-driven acceleration due to momentum conservation. However, this effect is strongly dependent on the density profile of the medium surrounding the outflow. If all the mass is concentrated on a thin spherical shell close to the source, then the mass-loading term is negligible after the initial acceleration and the wind propagates ballistically. In the opposite extreme, a uniform ambient density leads to an asymptotic behavior of $v\sim r^{-1/2}$, as described in Section \ref{sec:mass_size}. We adopt a power law profile for the density for the ambient medium (pre-existing gas or previous slow wind), $\rho(r)\propto r^{-\eta}$, so that we can decrease the effects of mass-loading by increasing the parameter $\eta$.

We use the parameters of the BN object to describe the central source. After fixing $\Gamma$ and a density profile, we can integrate eq. \eqref{eq:wind_dynamics} from $r_\text{min} = \SI{4}{R_\odot}$ up to $r_\text{max} = \SI{0.28}{pc}$ (the observed extent of the Orion outflow) to find the final velocity. Since the luminosities of the sources at the time the wind was initiated are uncertain, we use a range of Eddington ratios from just above 1 up to 2.5. We set the normalization of the density profile so that the wind sweeps up $\SI{1}{M_\odot}$, the approximate upper bound on the mass contained in the fingers.

\begin{figure}
    \centering
    \includegraphics[width=\linewidth]{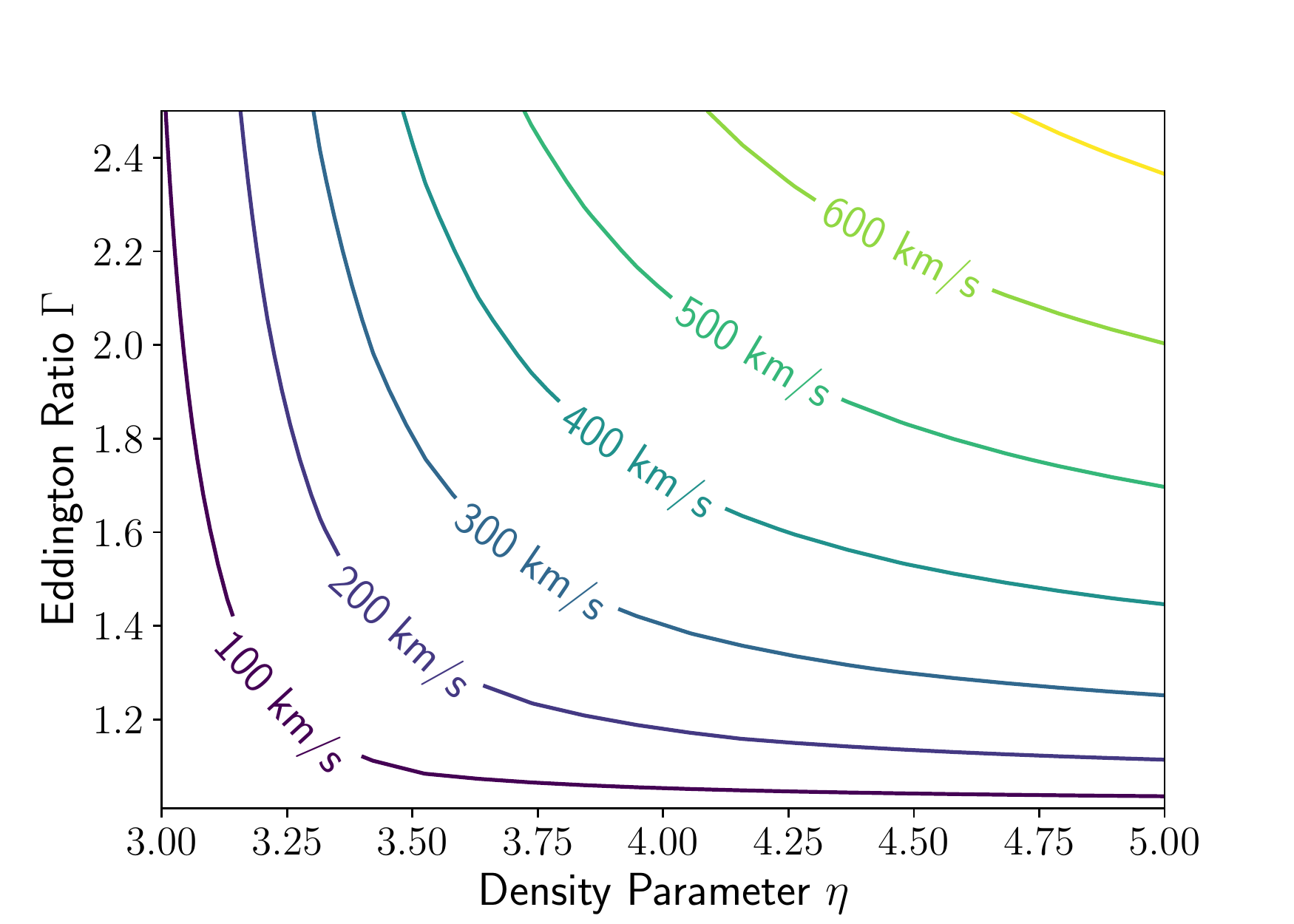}
    \caption{For selected values of the Eddington ratio $\Gamma$ and the density scale parameter $\eta$, we integrate eq. \eqref{eq:wind_dynamics} to obtain the wind speed at $r_\text{max} = \SI{0.28}{pc}$. Contours of the final speed are shown here. The necessary value of \SIrange{300}{400}{km/s} is achievable for a wide range of parameter values.}
    \label{fig:wind_speeds}
\end{figure}

Figure \ref{fig:wind_speeds} shows the wind speed $v_f$ reached at $r = \SI{0.28}{pc}$ as a function of the Eddington ratio and the density parameter $\eta$. There is a sizable region of parameter space for which $\SI{300}{km/s} < v_f < \SI{400}{km/s}$, which is what we expect for the wind which produced the Orion fingers. Generally, lower values of $\Gamma$ are more plausible. Values of $\eta$ are quite uncertain. The high values of $\eta$ (steeply declining density profile) suggested by this model may arise if the gas originated as a slow outflow from circumstellar disks before the final strong dynamical interaction. Lower values of $\eta$ are more plausible if the density follows that of the initial molecular cloud. Indeed, submm observations of proto-stellar cores with SCUBA-2 \citep{kirk05}, including observations of cores in the Orion molecular cloud \citep{kirk16}, suggest a nearly flat core within $\sim$\SI{0.05}{pc} of the center, followed by a roughly isothermal profile. In the following section, we show that when we relax the assumption of a thin spherical wind and simulate the hydrodynamics along with continual radiative driving, we can reach high velocities even with a density profile of this form.

\subsection{Constraints on mass and size of fingers}\label{sec:mass_size_wind}

\begin{figure}
    \centering
    \includegraphics[width=\linewidth]{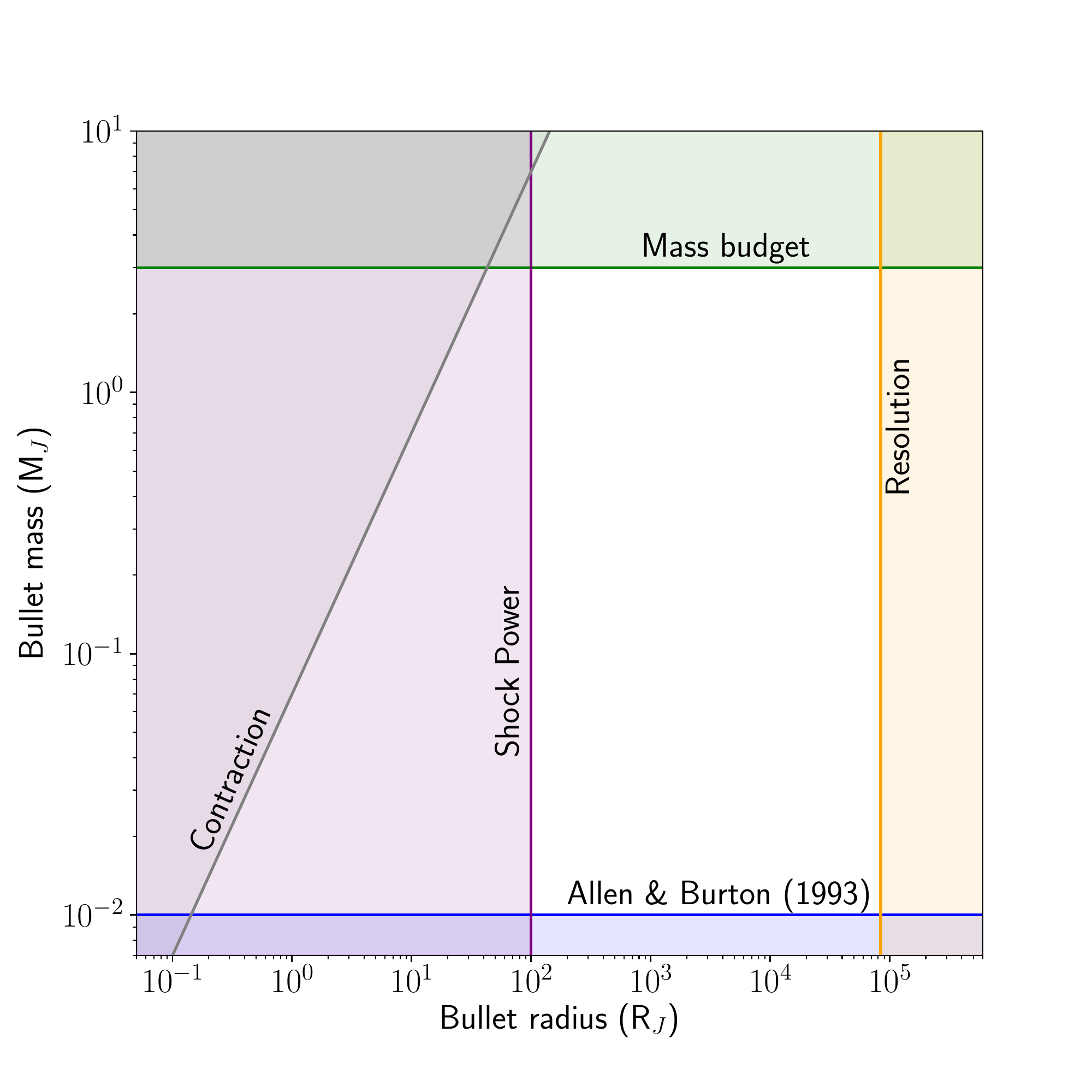}
    \caption{The various constraints on the mass and size of the Orion fingers, assuming they are formed \emph{in situ} via a hydrodynamical instability in a massive wind. These constraints are described explicitly in Section \ref{sec:mass_size_wind}.} 
    \label{fig:mr_diagram_wind}
\end{figure}

The basic observational constraints on the mass (the total mass budget, and the minimum mass required to account for iron lines), as well as the resolution limit of the Gemini telescope, are independent of the physical formation scenario. Since the dense clumps formed from the Rayleigh-Taylor instability are not undergoing Kelvin-Helmholtz heating, we expect them to be cooler than the bullets, and so we do not expect these clumps to be visible to Gemini as point sources. Additionally, the radius of the dense objects which have formed is still bounded below due to the limitations in cooling which can be achieved in $\sim\SI{e3}{yr}$, so \eqref{eq:contraction} still applies. Thus, these four constraints can be carried over directly from Section \ref{sec:mass_size}. They are shown in Figure \ref{fig:mr_diagram_wind} to illustrate the larger feasible region of parameter space.  However, the constraints which were specific to dense bullets propagating outwards from a central point -- especially the requirement that they be stable to tidal forces during their acceleration -- are no longer relevant. Instead we must consider the hydrodynamics relevant to the fragmentation of a stellar wind.

The formation of the bullets via a hydrodynamical instability is described by the interacting stellar wind model. In this model, a fast wind collides with a slower wind launched earlier, and the ensuing deceleration triggers a Rayleigh-Taylor instability which produces the Orion bullets. The fastest-growing modes of the RT instability are those with the smallest wavelengths that are not suppressed by dissipation effects. In molecular gas, the dissipation will be dominated by magnetic viscosity, though turbulence can also play a role \citep{krum15}. The fastest-growing mode in the linear regime sets an initial scale, but as nonlinear effects take over the growth of the instabilities becomes much harder to predict.

We can test the qualitative behavior of interacting stellar winds with numerical simulations. We use Athena++ \citep{whit16,felk18} to study the Rayleigh-Taylor instability. Below we describe one particular set-up, based on the hypothesis that the instability results from collisions of IR-driven winds. As we will show, these conditions are sufficient to produce Rayleigh-Taylor instabilities which result in a morphology similar to that seen in Orion. However, these conditions are not necessary. The same model in different regions of its parameter space, or rather different models such as that used in \cite{ston95}, also suffice to generate the Rayleigh-Taylor fingers.

The simulation runs on a 2D $512\times 512$ grid in polar coordinates, covering 1 radian and \SIrange{20}{3.4e4}{AU}. The initial density profile is taken to be proportional to $r^{-2}$, with the density on the inner boundary set to \SI{e8}{cm^{-3}} and the minimum initial density set to \SI{e4}{cm^{-3}}. Random pink noise proportional to $k^{-1}$, with a maximum angular frequency $k$ of 60 and a total amplitude of \num{e-2} relative to the mean density, is added to the inner \SI{2e3}{AU}. Previous simulations have used an isothermal equation of state, justified by a cooling time of \SI{1}{yr} compared to a dynamical timescale of $\sim\SI{1000}{yr}$ \citep{ston95}. We instead use an adiabatic equation of state with a sound speed of \SI{2}{km/s}, and explicitly include a cooling function with a timescale of \SI{1}{yr}. 

\begin{figure*}
    \centering
    \includegraphics[trim=6cm 4cm 0cm 4cm, clip,width=\linewidth]{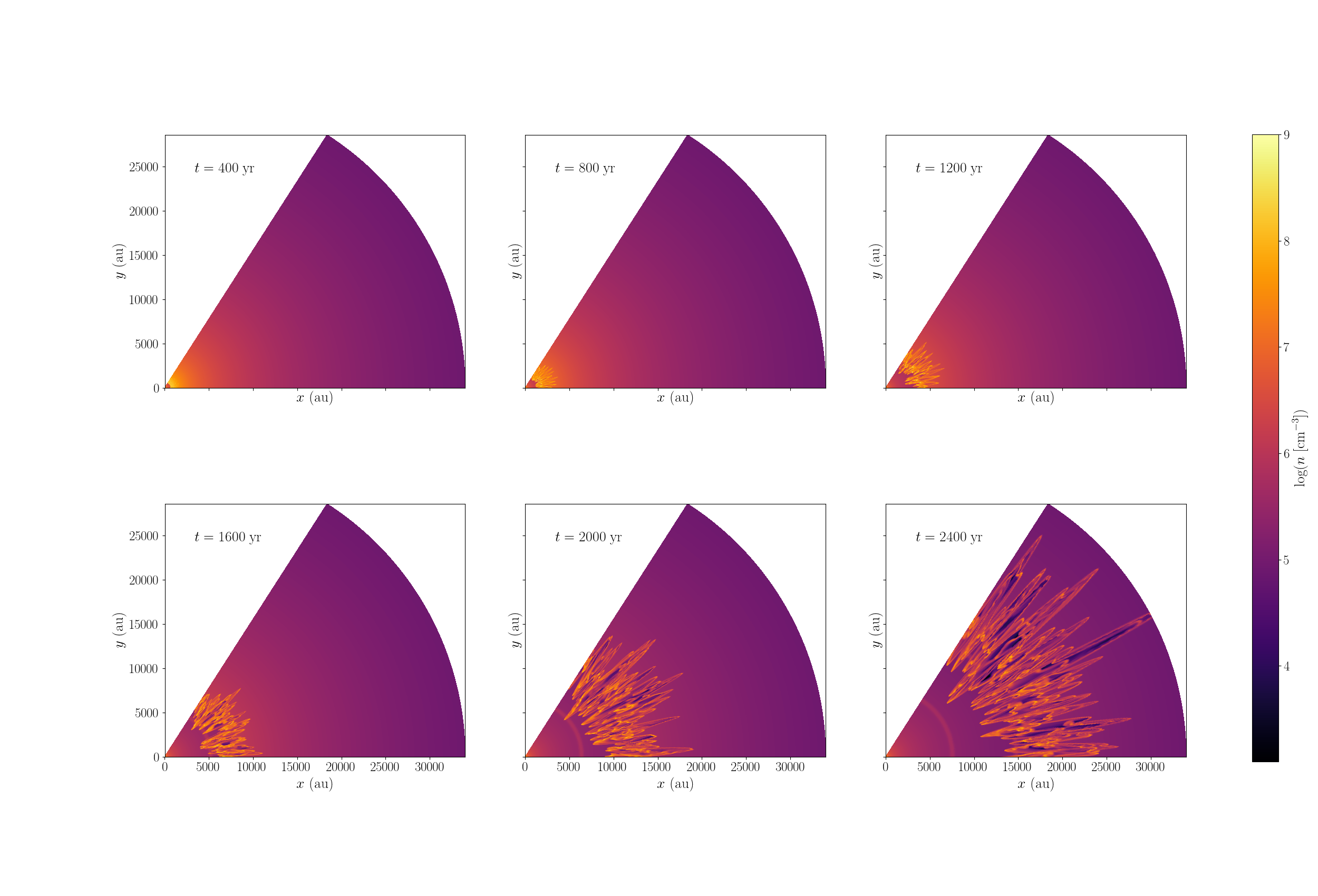}
    \caption{Six snapshots of hydrodynamic simulations using Athena++ \citep{felk18,whit16}. An IR driving source term is used to accelerate fast winds with collide with a growing slow wind, accelerating it and producing fingers via the Rayleigh-Taylor instability. Over the course of the simulation, the fingers reach velocities of several hundred km/s and exhibit a morphology similar to the observed Orion fingers.}
    \label{fig:athena}
\end{figure*}

Winds are launched via IR driving, making the simplifying assumption that photons are either transmitted or absorbed. We thus integrate the radiative transfer equation
\begin{equation}
    \d{I_\nu}{r} = -\kappa_\text{IR}\rho(r) I_\nu
\end{equation}
with $\kappa_\text{IR} = \SI{5}{cm^2/g}$ \citep{thom15,drai03} to compute the intensity of radiation throughout the domain at each time step, and from this compute the radiation pressure and the driving force at each point. Unlike equation (\ref{eq:wind_dynamics}), this equation does not take into account the momentum boost due to multiple scatterings and therefore is likely to underestimate the radiation pressure. We set the intensity at the inner boundary by placing a gravitational point source of \SI{8}{M_\odot} at the origin, and setting its Eddington ratio in the IR to 3. In order to produce fast winds which can interact with slow winds, we increase the luminosity by an order of magnitude for a duration of 10 years every 100 years.

Snapshots of the resulting density profile are shown in Figure \ref{fig:athena}. The fast winds driven out every 100 years repeatedly collide with a slower, denser wind, accelerating it and triggering Rayleigh-Taylor instabilities which form the finger morphology seen in Orion. We do not simulate magnetic fields, which play an important role in setting the Mach angles of the Orion fingers as discussed in Section \ref{sec:angles}, so we cannot directly compare the opening angles in our simulations to those observed in Orion. Nonetheless, Figure \ref{fig:athena} shows that the Rayleigh-Taylor instability is a plausible mechanism for forming the Orion fingers. Much like in the bullets scenario, the simultaneous launching of the fingers at a range of velocities naturally leads to a Hubble-type flow.

Generically, we expect that time-variable driving of a damping medium such as the ISM will lead to the onset of Rayleigh-Taylor instabilities. This is consistent with \cite{ston95}, who simulate a different variable wind model and find a similar resulting morphology. Our simulations are meant to demonstrate an example of how the instabilities might be formed, but there are many parameters which could vary without affecting the overall behavior. The behavior of the source luminosity in particular is motivated only by the need to inject sufficient energy into the system in a time-variable manner; the frequency or the energy scales involved could well be different without affecting the overall behavior. Likewise, the $r^{-2}$ density profile near the core is chosen to be consistent with submillimeter observations \citep{kirk05,kirk16}. We have tested the simulation for density profiles ranging from $r^{-2}$ to $r^{-4}$ and shown that the qualitative behavior is unchanged, suggesting that the steep density profile in the previous section is an artifact of the thin-shell wind model.

\subsection{Feasibility}\label{sec:wind_feasibility}

The hydrodynamical instability scenario has several advantages relative to the accelerated bullet scenario of Section \ref{sec:bullets}. We have already discussed at some length one of the greatest advantages: young stellar winds are naturally capable of producing velocity scales commensurate with those of the observed Orion bullets, whereas such velocities are difficult to reach via gravitational acceleration of bullets. No significant fine tuning is required to accelerate a wind to hundreds of kilometers per second, as shown in Figure \ref{fig:wind_speeds}. Furthermore, these fast winds can readily fragment and form a morphology similar to what is observed in the Orion outflow, as shown by our simulations and others \citep{ston95}.

In addition to the $\text{H}_2$ fingers, there is a natural explanation for the formation of the CO streamers in the context of this scenario (Bally, priv. comm.). As the Rayleigh-Taylor instability develops, there will be shear along the edges of the fingers, leading to the development of Kelvin-Helmholtz instabilities. This will lead to dense clumps moving in the same direction as, but possibly slower than, the fingers. These clumps would thus have the kinematics necessary to be the progenitors of the CO streamers, which trail behind the $\text{H}_2$ fingers \citep{ball17}. 

Furthermore, the physical characteristics of the propagating objects are less constrained in this scenario. In the first scenario, in order for bullets to survive acceleration to hundreds of kilometers per second, they must fall into a relatively narrow range of masses and sizes which will keep them stable against the large tidal forces they experience, either via self gravity or ram pressure confinement, and furthermore they must satisfy observational constraints. This is not a concern if the Orion fingers formed out of a wind, since they never would have experienced significant tidal forces. There is plenty of configuration space available for fingers which condensed out of a wind and which satisfy all present observational constraints, as shown in Figure \ref{fig:mr_diagram_wind}.

\begin{figure}
    \centering
    \includegraphics[width=\linewidth]{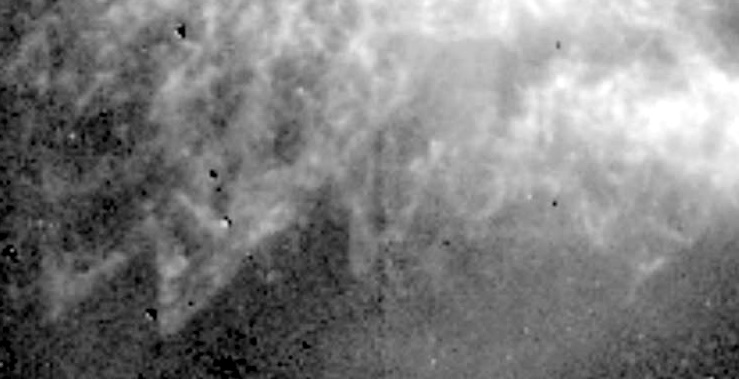}
    \caption{Part of the northeastern edge of the nebula M1-67 surrounding WR 124 in the H$\alpha$ filter. These bow shocks are observed throughout the nebula, although they are most clear on the edges where they are unobscured. They are believed to be the result of Rayleigh-Taylor instabilities in highly variable stellar winds, where new fast wind is interacting with the material from previous ejections \citep{gros01}. This image has been made available by NASA and STScI, and contrast-adjusted by the authors to highlight the shocks.}
    \label{fig:wr}
\end{figure}

However, one key question is left open in the hydrodynamical instability scenario: why is the outflow coeval with the stellar interaction? In fact, it is not clear that the dynamical interaction of BN, Source I, and source x is necessary for a stellar wind several thousand AU away to fragment. A phenomenon similar to the Orion bullets has been observed in the ejection nebula M1-67 surrounding the Wolf-Rayet star WR 124. Figure \ref{fig:wr} shows a section of the northeastern edge of the nebula, where some bow shocks are visible. This is believed to be the result of Rayleigh-Taylor instabilities in interacting stellar winds \citep{gros01}, which shows that a single star with a radiatively driven wind is sufficient to produce the morphology quite similar to that of the Orion fingers. 

Therefore, the dynamical interaction between the stars may not, strictly speaking, be necessary for the formation of a handful of fingers. However, it is hard to imagine that the spectacular BN/KL outflow is coeval with the dynamical interaction of the stars by sheer coincidence. Instead, we consider how the dynamical interaction may have triggered or enhanced the formation of the Orion fingers, and in particular enhance the available mass supply and the hydrodynamical instabilities.  

\cite{mcca97} suggest that the variability in the stellar wind which triggers the Rayleigh-Taylor instability is caused by the stars becoming nearly coincident during their interaction and driving the same shell. This would naturally explain why the fingers have kinematic timescales between 500 and 2000 years, suggesting that they were launched during the stellar interaction. Indeed, our simulations suggest that the speed of the fingers scales with the speed of the fast wind which formed them, so if the fast wind were launched during the stellar interaction, we would expect to see fingers with similar kinematic timescales.

However, given the disparity in the masses of the stars, and the strong dependence of luminosity on mass, it is unlikely that the combination of two winds would represent a significant variability of the stronger wind. The interaction of the stars might instead be responsible for providing additional mass. For instance, if one of the stars had already driven out a slow wind before the interaction, then it would be surrounded by a shell of swept-up material. As another star approaches the first star during the interaction, its circumstellar material would be disrupted from orbit and could be accelerated as a second, faster wind. The steep density profile in this second wind would itself be conducive to RT instabilities, and moreover, the deceleration upon impact with the first wind would trigger the formation of fingers like those in Figure \ref{fig:athena}. Alternatively, the hypothetical merger of stars I$_1$ and I$_2$ into Source I \citep{fari18} could have resulted in an expulsion of some part of the stellar envelope, providing material for the second wind \citep{ball16}. These are just some of the possibilities for how the interaction of the stars results in a greater supply of mass for ejection. 

Furthermore, we hypothesize that finger systems resulting from stellar interactions are more likely to be observed than ones driven by single stars. In Section \ref{sec:bullet_formation}, we showed that ram pressure confinement lasts longer for more massive objects. Thus, unless the hydrodynamical instability forms objects compact enough to be self-gravitating, the time for which the finger morphology persists is directly related to the amount of mass available. For a single star to sweep up enough mass in a wind to form fingers which last for a significant amount of time, it must be extremely massive and luminous itself (as is WR 124, with a mass of \SI{33}{M_\odot} and a luminosity \SI{e6}{L_\odot}, \citealt{hama06}). If instead multiple stars have their circumstellar material combined into the same spherical outflow, which can readily occur in the potential well of a star-forming nebula, sufficient mass for the formation of fingers is more easily available.

\section{Discussion}\label{sec:disc}

\subsection{Opening angles of the bow shocks}\label{sec:angles}

Opening angles of the wakes behind the bullets provide additional constraints on the physical conditions in the wind region. We assume that the timescales for formation and acceleration of the bullets are much shorter than the age of the system. As the first approximation, we assume that all bullets are launched simultaneously with a range of velocities ${\bf v_i}$ and propagate over duration $\tau$ ballistically from the launching site (similar to the model of \citealt{ball11}). If the effective sound speed in the surrounding medium is $c_s$, then the physical half-opening angles of the Mach cones of the wakes are $\sin \alpha_i=c_s/v_i$. As seen in projection on the plane of the sky, for a bullet propagating at angle $\theta$ relative to the line of sight the observer detects the bullet at the projected distance $v_i \tau \sin\theta$, with an apparent opening half-angle $\tan\alpha'_i=\tan\alpha_i/\sin\theta$. If the bullets are highly super-sonic, with $\tan \alpha \simeq \sin\alpha$, then we expect that the projected distances $d_i$ and projected opening half-angles $\alpha'_i$ are related by 
\begin{equation}
\sin \alpha'_i\simeq c_s \tau / d_i. 
\label{eq:projected}
\end{equation}
In forming a single locus on the $\alpha'$ vs $d$ space despite a potentially large range of velocities within the outflow, this relationship is reminiscent of the relationship between the proper motions and the projected distances for a coeval outflow investigated by \cite{ball11}.

\begin{figure}
    \centering
    \includegraphics[width=.9\linewidth]{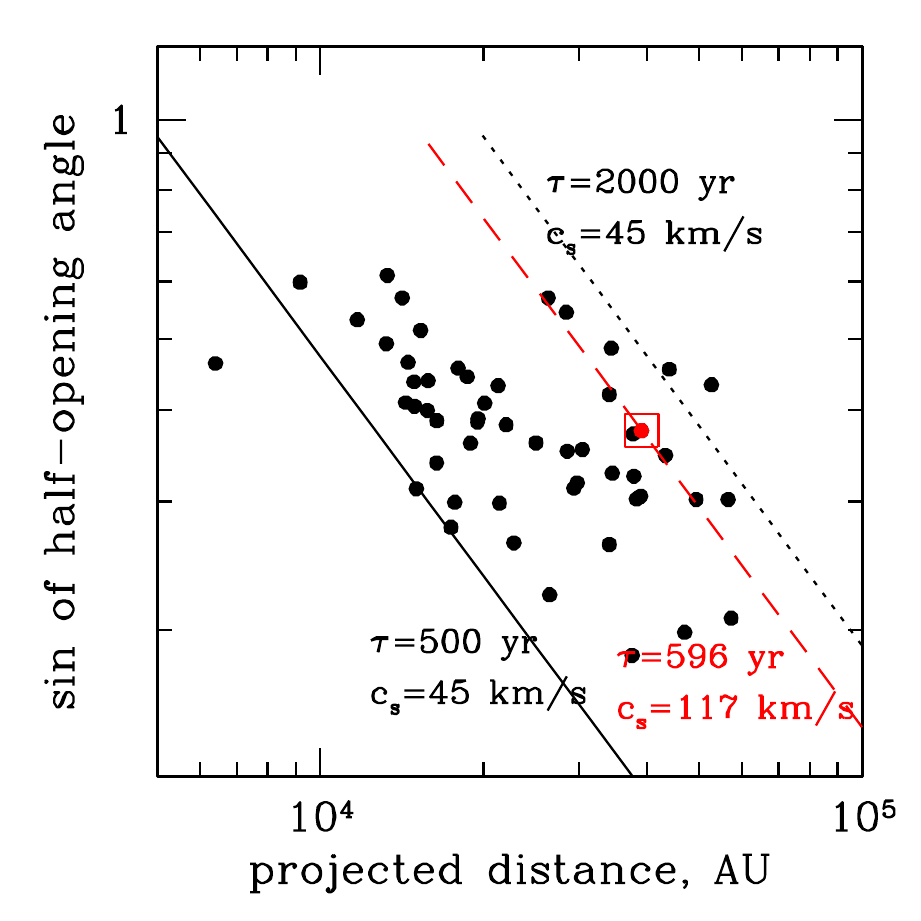}
    \caption{The distribution of projected distances and projected opening angles for 53 well-defined, conically shaped wakes. The inclined lines show equation (\ref{eq:projected}) with parameters as indicated. The point highlighted with a red square corresponds to the bullet with an accurate proper motion measurement and a corresponding kinematic age of 596 years \citep{ball15}. We find that for a plausible range of ages, the required effective sound speeds are too high for any medium with $T\la 10^4$ K.}
    \label{fig:opening_angles}
\end{figure}

We use the color-composite image from \cite{ball15} to calculate the apparent half-opening angles of 53 well-defined (conically-shaped and isolated) wakes by visually identifying the vertex and points on the sides and recording their pixel positions. These measurements are shown as a function of the projected distance (in AU, calibrated from the pixel units of the color composite to the physical units using the positions of several stars in the field) in Figure \ref{fig:opening_angles}. In general the observed points do follow the $\sin \alpha'_i \propto 1/d_i$ trend, but with a large spread, which can be due to the spread in $c_s$, $\tau$ or both. The minimal age of the outflow can be determined from the proper motions of the runaway stars to be 512 years \citep{ball11}. We place the corresponding lower bound on the observed $\alpha'_i$ with a solid black line using $\tau=500$ years and $c_s=45$ km/s (the one outlying point may be an unrelated Herbig-Haro object). Fixing this sound speed and considering the range of ages suggested by the position-velocity diagram \citep{ball11}, we place the dotted line at $\tau=2000$ years as the upper envelope of the data. 

By volume, much of the gas into which the Orion fingers' bow shocks propagate must be molecular, since we directly see H$_2$ and CO line emission in the wakes and the molecular formation time is much too long for these molecules to form as the shocks propagate into the gas \citep{glov10}. Yet the sound speed estimated above is much higher than the one expected in molecular gas at several hundred K (a few km/s), or the one in photo-ionized gas at $10^4$ K ($\sim 20$ km/s). To disentangle the age and the sound speed parameters in equation \eqref{eq:projected}, we cross-correlate our list of geometrically measured wakes with the list of bullets for which \cite{ball15} measure proper motions and we find one object in common (highlighted with a red square), with a kinematic age -- projected distance divided by the proper motion -- of 596 years. With this age and the apparent half-opening angle of 22$^\circ$, this bullet would need to be propagating through the medium with an effective sound speed $c_s=117$ km/s, shown with the dashed red line. 

In other words, the observed opening angles of the cones are much larger than one would expect for a warm ionized medium, let alone for a diffuse molecular medium, given that \SIrange{300}{400}{km/s} projected velocities have been directly detected in the outflow using proper motions \citep{ball11}. Indeed, if the thermal sound speed in the gas really is \SI{45}{km/s} or higher, the gas would have a temperature on the scale of megakelvins and we would expect to see diffuse X-ray emission from the interiors of the fingers. The Chandra Orion Ultradeep Project conducted an extensive X-ray survey of the OMC-1 region, and did detect X-ray emission from the tip of the finger with the highest proper motion, but X-rays were not detected from any of the other fingers \citep{getman05,grosso06}. There must be some other explanation for the large opening angles.

One possibility is that the effective sound speed is affected by the magnetic fields. Magnetosonic waves at an angle $\theta$ with the constant background field $\bm{B}_0$ have a double-branched dispersion relation with wave speeds
\begin{align}
    \begin{split}
        v_0 &= v_A\cos\theta,\\
        v_\pm^2 &= \frac{1}{2}\left(v_A^2 + \overline{c_s}^2 \pm \sqrt{v_A^4 - 2v_A^2 \overline{c_s}^2 \cos(2\theta) + \overline{c_s}^4}\right).
    \end{split}
\end{align}
Here, $v_A^2 = B_0^2/4\pi\rho_0$ is the Alfv\'en velocity in the background density $\rho_0$, and $\overline{c_s}$ is the sound speed in the zero-magnetic field limit. The speeds satisfy $v_+ > v_0 > v_-$, so the Mach cone opening angle is set by $v_+$. The maximum value of $v_+$ is $v_\text{max} = \sqrt{v_A^2+\overline{c_s}^2}$, which occurs when $\theta = \pi/2$. Thus, the maximum half-opening angle for a Mach cone is
\begin{equation}\label{eq:opening_angle}
    \alpha = \sin^{-1}\frac{v_\text{max}}{v},
\end{equation}
generated by a source moving at velocity $v$ transverse to the background magnetic field.

Figure \ref{fig:opening_angles} shows that $v_\text{max}$ must be at least $\SI{45}{km/s}$ in order to be consistent with the opening angles of the conical wakes of the Orion fingers. With a background density of $\rho_\text{ISM} = \SI{e4}{cm^{-3}}$, this requires $B_0 \ga \SI{2}{mG}$, just above the value \SI{1}{mG} measured in the outflow on large scales using far-infrared polarimetry \citep{chus19}. Shock compression of the magnetic field can increase the effective Alfv\'en velocity relative to what we would expect from the measured average field.

These magnetic field values are strikingly high for molecular gas at a few hundred Kelvin. The ratio of magnetic to thermal energy densities is
\begin{equation}
    \frac{\rho_B}{\rho_T} = 190 \left(\frac{B_0}{\SI{1}{mG}}\right)^2 \left(\frac{n}{\SI{e4}{cm^{-3}}}\right)^{-1}\left(\frac{T}{\SI{e2}{K}}\right)^{-1},
\end{equation}
so clearly the magnetic fields are not in equipartition with thermal energy of the ISM. Such magnetic fields would be closer to equipartition (although still a factor of a few above equipartition) for the supersonic turbulence with $v_{\rm turb}=5-10$ km sec$^{-1}$ seen in Orion in ionized and in molecular gas \citep{arth16, orki17}, although it is not clear whether these measurements (made in different parts of Orion) apply to the particular wind environment in this study where presumably the turbulence is currently being generated by the outflow itself. The large opening angles of the bow shocks are somewhat unexpected and provide independent support for field values on the order measured by \cite{chus19}. \cite{chus19} argue that these high magnetic fields are generated as a result of the outflow, but the physical origin of the fields is not well understood. 


\subsection{Morphology}\label{sec:disintegration}

\begin{figure}
    \centering
    \includegraphics[width=\linewidth]{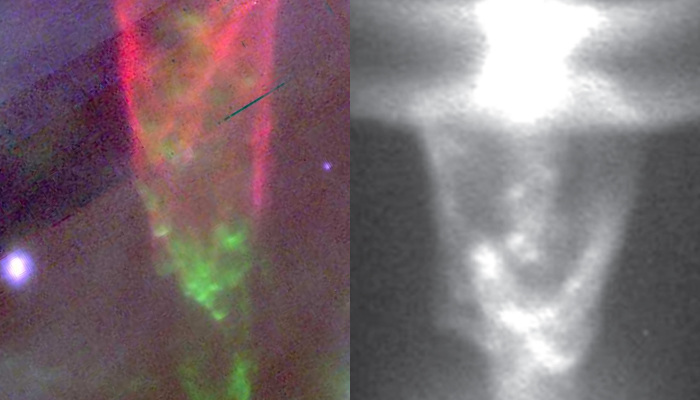}
    \caption{(Left) One of the Orion fingers, in which the shock head shows apparent fragmentation. Reprinted with permission from Figure 1 of ``The Orion fingers: Near-IR adaptive optics imaging of an explosive protostellar outflow'' \citep{ball15}. (Right) A laboratory bow shock, showing similar morphology due to thermal instability. Reprinted with permission from Figure 2 of ``Bow Shock Fragmentation Driven by a Thermal Instability in Laboratory Astrophysics Experiments'' \citep{suzu15}.}
    \label{fig:fragmentation}
\end{figure}

The Gemini image of the outflow shows that some of the fingers appear to be fragmenting, such as in the left panel of Figure \ref{fig:fragmentation}. This presents a challenge to both formation scenarios we have outlined. 

If the Orion fingers come from dense bullets which have propagated roughly 500 times their diameters \citep{ball15}, it is difficult to explain why they would just now be breaking up. It is possible to explain this by treating the bullets as ram pressure confined, and considering the difference in ram pressure acting on a bullet propagating into ISM versus a bullet propagating into the post-shock material created by a bullet ahead of it. However, this explanation requires that the bullets lie in an extremely narrow range of their parameter space.

If the Orion fingers are generated by structures which formed \emph{in situ} in a stellar wind, then the fragmentation is less surprising, but still curious. We would expect that whatever structures formed due to a hydrodynamic instability are either stable, such that they generate a clean bow shock, or unstable, such that they dissipate before forming a shock at all. The observations seem to suggest that instead the structures live on the edge of stability, lasting long enough to form a shock but then breaking up.

We propose to explain these morphological features by thermal instability. \cite{suzu15} have shown experimentally that thermal instability in bow shocks leads to a fragmentary morphology similar to what we observe in the OMC-1 outflow. They use plasma jets with radius $r_\text{exp} = \SI{1.5}{mm}$ and velocity $v_\text{exp} = \SI{140}{km/s}$, and observe fragmentation at $t_\text{exp}\sim\SI{150}{ns}$. Their experiments are constructed to have similar dimensionless parameters to those found in outflows around young stellar objects, including the cooling behavior, so we can use Eulerian scaling relations to determine the time scale for fragmentation in the Orion system \citep{ryut99}. The time $\tau$ for fragmentation of the Orion fingers is given by
\begin{equation}
    \tau = t_\text{exp}\frac{R_f}{r_\text{exp}} \frac{v_\text{exp}}{v_b},
\end{equation}
where $R_f$ is the size of the shock heads (which we expect to be larger than the size of the bullets), and $v_b = \SI{300}{km/s}$ is the typical velocity of outer fingers which show the fragmentation. From the Gemini images of the outflow, we estimate $R_f$ to be \SIrange{50}{150}{AU}. This gives $\tau$ on the order of 100 years, which is the order of magnitude needed to explain the beginnings of fragmentation in the Orion system.


\subsection{Future fate of the Orion bullets}\label{sec:fate}

The kinematics of the Orion fingers, as well as the stars in the vicinity of the OMC-1 outflow, strongly suggest that the key event which led to the formation of the fingers took place 500--1000 years ago. The present spectacular outflow is a transient phenomenon which will eventually disappear as the bullets recede further from their point of origin and decelerate. Additionally, the bow shocks themselves may fragment and dissipate further as they succumb to the thermal instabilities described in \cite{suzu15}. However, the fate of the progenitors of the shocks, small dense objects which are presently moving supersonically through the interstellar medium, is unclear. If they are stable and survive in the long term as self-gravitating bodies, then OMC-1 is presently undergoing a planet formation process entirely different from the standard situation of bodies forming out of circumstellar disk material while in orbit around a star. Given that the system in Orion appears not to be unique in the galaxy \citep{ball16}, this process could be responsible for the formation of a sizeable population of free-floating planets.

In our first scenario, described in Section \ref{sec:bullets}, the initial formation of the bullets looks quite similar to one route for planet formation \citep{boss97}: gravitational instabilities in a disk collapse and form dense clumps of gas, which continue contracting into planets. After this stage, however, the bullets participate in dynamical interaction of the BN object, Source I, and source x. Only about 1000 years elapse between the beginning of the interaction triggering gravitational instabilities and the interaction itself. On this short timescale, the bullets have to become stable enough to survive the large tidal forces which accompany their acceleration to hundreds of kilometers per second.

If self gravity kept the bullets stable during the initial acceleration, then they would remain stable now, and they would form free-floating planets. If the bullets were in the regime where they were initially held together by ram pressure, then their fate depends on how much they can contract before ram pressure deconfinement. From equation (\ref{eq:deconfinement_time}), it follows that bullets which satisfy condition (\ref{eq:ram_upper}) will be confined for at least $\sim\SI{3000}{yr}$, and longer if they have greater than the minimum mass. At \SI{e-2}{M_J}, the cooling times are too long to allow the object to contract sufficiently in this length of time. However, for a mass on the order of \SIrange{0.1}{1}{M_J}, the confinement time is longer and the contraction time is shorter, making it plausible that the bullet reaches a core temperature of \SI{2500}{K} and undergoes rapid collapse due to hydrogen dissociation \citep{naya15}. Thus, in our first scenario, it is overall relatively likely that the more massive of the Orion bullets, will form free-floating planets.

In the second scenario, described in Section \ref{sec:winds}, the Orion bullets condensed out of a stellar wind following the onset of a hydrodynamical instability. In this case, we have less of a clear grasp on the mass and size of the bullets. However, the timescale for ram pressure deconfinement remains the same, modulo an additional 500 years of uncertainty since we do not know when the bullets formed in this case. For bullets in the low mass, large size section of the feasible region in Figure \ref{fig:mr_diagram_wind}, the deconfinement time will be only a few hundred years, not long enough for the bullets to contract significantly and become self-gravitating. However, if the bullets are more massive and smaller, such that the deconfinement time reaches \SIrange{e4}{e5}{yr}, then like in the first scenario, there may be sufficient time for the bullets to reach \SI{2500}{K} and collapse rapidly. Thus, in this scenario, it is unclear whether the Orion fingers will spawn a long-lasting population of free-floating planets, or whether the phenomenon will dissipate completely.

We would be in a better position to assess the long-term fate of hydrodynamically formed bullets if we had a better idea of their likely masses and sizes. This is difficult due to the inherent uncertainty in the history of the winds from which they formed. Our simulations in Section \ref{sec:mass_size_wind} qualitatively represent how a Rayleigh-Taylor instability might form dense objects moving at high velocities and forming bow shocks, but we do not explore anything close to the full parameter space for the initial conditions or the driving mechanism. A comprehensive set of simulations might shed more light on the likely parameters of hydrodynamically formed bullets, but we leave this for future work.

If the Orion bullets do survive as free-floating planets, we expect for them to be physically and chemically distinct from ordinary Jovian planets. The short formation times imposed by the timescale of the stellar interaction (\SI{e3}{yr}) would leave little time for the formation of an icy core \citep{hell08}, although core formation on this timescale may be possible in some rare cases \citep{naya15}. Thus, the resulting planets would likely be composed almost entirely of gas. Without a solid phase, chemical differentiation requires far too much time even at the minimum possible temperatures \citep{mccr65}, and the Orion bullets would be better mixed than typical planets. These features would distinguish free-floating planets originating in Orion-type outflows with those that are ejected long after they are formed during dynamical interactions with other planets orbiting the same star \citep{juri08}. Furthermore, the Orion bullets have much greater velocities than what could be generated by planet-planet dynamical interactions.

A completely independent constraint on the formation of free-floating planets comes from gravitational microlensing. In principle, microlensing is the ideal way to survey for unbound planet-mass objects, since the detection does not depend on the luminosity of the object and the mass can be inferred from the duration of the lensing event \citep{mroz17}. However, the estimated number of free-floating planets in the galaxy has varied significantly between different studies. It was originally suggested that there may be as many as two Jupiter-mass objects per main sequence star \citep{sumi11}. It was originally suggested that there may be as many as two Jupiter-mass objects per main sequence star \citep{sumi11}. These estimates assume typical Galactic velocities ($v_{\rm assumed}\sim 120$ km sec$^{-1}$ for the Bulge; \citealt{sumi11}), so if the free-floating planets have higher velocities $v_{\rm actual}=300$ km sec${-1}$ characteristic of the Orion outflow, then the $\sim 1 M_{\rm J}$ masses inferred by \citet{sumi11} would need to be revised upward by a factor of $v^2_{\rm actual}/v^2_{\rm assumed}\simeq 6$. More recent measurements with a larger sample of microlensing events have found an upper bound of 0.25 of these objects per main sequence star \citep{mroz17}, but also indicated an excess of ultra-short events, which could be either due to Earth-like planets with normal disk or bulge kinematics or to more massive planets with higher velocities. Whether objects like Orion bullets form a major or a minor fraction of this population depends critically on whether the bullets are currently or eventually self-gravitating, which is the question we started to address in this paper but have not yet fully resolved.

\section{Conclusion}\label{sec:conc}

``Orion fingers" is a spectacular system of dozens of bowshocks, with shock-heads (``bullets") propagating with velocities of up to 300 km/s away from a common center of origin, where they apparently originated 500-1000 years ago. The center of this outflow is also a suspected site of a strong dynamical interaction of a $\sim4$-body system of massive young stars, which are now escaping from the region with velocities 10-50 km/s. In this paper we examine the nature of the bullets, discuss constraints on their sizes and mass presented in Figures \ref{fig:mr_diagram} and \ref{fig:mr_diagram_wind}, and investigate two plausible scenarios for the formation of the ``fingers" with a focus on explaining the observed properties of the system without fine-tuning.  

In the first scenario, massive young stars with marginally stable, gas-rich disks find themselves in a strong dynamical interaction, which is common in cores of star-forming clouds. The gravitational perturbations of the disks during the interaction induce a period of rapid ($\sim 1000$ years) planet formation. The resulting planets, which form mostly by gravitational collapse rather than accretion of an envelope onto an icy core, are then ejected from the system during the same close passage of the massive stars that results in the dynamical disintegration of the stellar system. This scenario naturally explains the close timing coincidence between the origin of the outflow and the escape of the massive stars and predicts that the resulting bullets are self-gravitating planets with masses between 0.01$-$1 $M_J$. 

The weak point of this scenario is the difficulty of reproducing the velocity distribution of the escaping planets. Since the distance of the closest approach for the stars is well-constrained by the velocities of the ejected stars to be $\sim 0.5$ AU, the cross-section for ejecting planets at 300 km/s (which would require a passage well within 0.1 AU) is small. However, as our calculations do not include gas-dynamical effects, it is possible that funneling the gas toward the stars during the interaction might enhance the cross-section. Whether planets can form quickly enough and be ejected fast enough should be explored with more detailed gas-dynamical simulations. 

In the second scenario, the gas disks around the young stars are disrupted during the interaction, and the radiation pressure of the massive stars on the resulting gas cloud initiates a massive, likely infrared-driven, stellar wind, which then fragments into clouds due to Rayleigh-Taylor and other instabilities. This scenario naturally explains the high velocities of the bullets, as hundreds of km/s is a common velocity for radiatively driven winds from massive stars. 

The timing coincidence between the stellar interaction and the outflows in this scenario is explained by the greater availability of gas during the stellar interaction and a better coupling between the stellar radiation and the surrounding gas than could be achieved in a quiescent disk. However, the fate of the bullets in this scenario is not known. If the instability forms objects of high enough mass, the cooling time is short enough that they become self-gravitating during the interval in which they are hydrodynamically confined.

While Orion fingers are visually spectacular, they are not necessarily unique in the Galaxy \citep{ball16}. Several examples of similar morphology are already known \citep{saha08, zapa13} and more may be uncovered in surveys of shocked emission around young stellar objects \citep{eisl00, gute04}, especially via searches for excess 4.5 \micron\ emission due to shocked molecular hydrogen \citep{cyga08}. At least one of the scenarios we discuss in this paper is likely to be common to star-forming regions and may result in a formation of free-floating planets with very high velocities (dozens to hundreds km/s) and with structural properties quite different from giant planets formed via core accretion. 

\acknowledgments

Simulations in this paper made use of the REBOUND code which can be downloaded freely at \url{http://github.com/hannorein/rebound}, as well as Athena++ which is available at \url{https://github.com/PrincetonUniversity/athena-public-version}. We thank the authors of \cite{ball15} and \cite{suzu15} for permission to reprint their figures in this work. We thank the referee J. Bally for many interesting suggestions, as well as acknowledge useful conversations with him prior to the submission of the paper. N.L.Z. acknowledges useful conversations with M. Begelman, V. Beskin, J. Krolik, A. Sternberg, T. Thompson and S. Tremaine. N.L.Z. further acknowledges the generous support by the Deborah Lunder and Alan Ezekowitz Founders' Circle Membership at the Institute for Advanced Study where this work was started. 

\bibliography{yso}{}
\bibliographystyle{aasjournal}

\end{document}